\documentclass[12pt]{article}
\usepackage{epsfig}
\usepackage{psfrag}
\usepackage{latexsym}
\usepackage{indentfirst}
\usepackage{fancyhdr}
\usepackage{amssymb}
\usepackage{amsmath}
\usepackage{amsfonts}
\usepackage{pifont}
\usepackage{cite}
\usepackage{color}
\usepackage{slashed}
\usepackage{graphics}
\usepackage{url}
\usepackage{array}
\usepackage{graphicx}
\usepackage{subfigure}

\def\be{\begin{equation}}
\def\ee{\end{equation}}
\def\bc{\begin{center}}
\def\ec{\end{center}}
\def\bea{\begin{eqnarray}}
\def\eea{\end{eqnarray}}
\newcommand{\bi}{\begin{itemize}}
\newcommand{\ei}{\end{itemize}}                 

\newcommand{\ba}{\begin{array}{c}}
\newcommand{\bad}{\begin{array}{ccc}}
\newcommand{\ea}{\end{array}}

\newcommand{\dr}{{\rm d}}
\def\nn{\nonumber}

\begin{document}

\begin{titlepage}
 \begin{flushright}
RM3-TH/14-13
 
 NORDITA-2014-89
 \end{flushright}
% % 
   \vskip 1cm
   \begin{center}
    {\Large\bf Effects of intermediate scales on renormalization group running of fermion observables in an SO(10) model}
   
   \vskip 0.2  cm
   \vskip 0.5  cm
Davide Meloni$^{\,a,}$\footnote{E-mail: \texttt{meloni@fis.uniroma3.it}},
Tommy Ohlsson$^{\,b,}$\footnote{E-mail: \texttt{tohlsson@kth.se}},
Stella Riad$^{\,b,}$\footnote{E-mail: \texttt{sriad@kth.se}}
\\[1mm]
   \vskip 0.7cm
% % 
% % 
% % 
 \end{center}

\centerline{$^{a}$ \it  Dipartimento di Matematica e Fisica, 
Universit\`{a} di Roma Tre,}
\centerline{\it  Via della Vasca Navale 84, 00146 Rome, Italy} 
\vspace*{0.2cm}
\centerline{$^{b}$ \it Department of Theoretical Physics, School of Engineering Sciences,}
\centerline{\it KTH Royal Institute of Technology -- AlbaNova University Center,}
\centerline{\it Roslagstullsbacken 21, 106 91 Stockholm, Sweden}
\vspace*{1.5cm}

\begin{abstract}
\noindent
In the context of non-supersymmetric SO(10) models, we analyze the renormalization group equations for the fermions 
(including neutrinos) from the GUT energy scale down to the electroweak energy scale, explicitly taking into account the effects 
of an intermediate energy scale induced by a Pati--Salam gauge group. To determine the renormalization group running, we use a 
numerical minimization procedure based on a nested sampling algorithm that randomly generates the values of 19 model parameters 
at the GUT scale, evolves them, and finally constructs the values of the physical observables and compares them to the existing 
experimental data at the electroweak scale. We show that the evolved fermion 
masses and mixings present sizable deviations from the values obtained without including the effects of the intermediate scale.

\end{abstract}
\end{titlepage}

\section{Introduction}
The lack of signals from physics beyond the Standard Model (SM) at the Large Hadron Collider (LHC) revives the question of 
which model constitutes the most appropriate extension of the SM and, if there is one, what is the energy scale where new 
features of particle interactions ought to be observed. The failure of the criterion of naturalness for new physics has caused 
a renaissance for models which aim to accommodate as much of the present state of knowledge as possible, while ignoring the 
fine-tuning problem \cite{Bajc:2005zf,Altarelli:2013aqa}. In the construction of a realistic model beyond the SM, one is, 
in principle, free to choose what features to be considered important. However, it is usually common practice that any new
model should, at least, contain a unification scale compatible with a naive expectation for the proton life-time as well as a 
Yukawa sector compatible with low-energy data. In addition, the model should allow for accommodation of a dark matter candidate as 
well as the baryon asymmetry of the Universe.

In the present work, we will study a non-supersymmetric extension of the SM model based on the gauge group SO(10), 
which has often been discussed in the previous literature 
\cite{delAguila:1980at,Harvey:1981hk,Robinett:1982tg,Mohapatra:1982tc,Babu:1992ia,Deshpande:1992au,Matsuda:2000zp,
Matsuda:2001bg,Bertolini:2009qj,Bertolini:2012im,Buccella:2012kc}.
The gauge group SO(10) has the clear advantage that all SM fermions, including right-handed neutrinos, 
belong to the same $16$ representation. However, the realization of the mechanism for the breaking to the SM gauge 
group requires the presence of large Higgs representations, and the consequent split of the multiplets to mass ranges 
differing in orders of magnitudes is an issue which, so far, has no satisfactory solutions in non-supersymmetric scenarios. 
Ad-hoc assumptions have been introduced \cite{delAguila:1980at,Mohapatra:1982tc}, which allow for the choice of the multiplets 
of the Higgs representations taking part in the evolution of the coupling constants. In particular, if a member of a Higgs multiplet has a vacuum expectation value (vev), $v$, 
corresponding to the breaking of a subgroup, then the mass of the whole multiplet is ${\cal O}(v)$ and will thus not 
contribute to the evolution of the coupling constant for energies below $v$, whereas for energies above $v$, the multiplet will have a mass of the order of the next, larger, mass scale where the larger symmetry appears.

In general, the viability of an SO(10) model is based on the ability to reproduce the values of fermion masses and mixings at the 
electroweak (EW) scale, $M_{\rm Z}$. Recent fits to fermion observables in non-supersymmetric contexts, which are discussed in 
Refs.~\cite{Joshipura:2011nn,Altarelli:2013aqa,Dueck:2013gca}, show that a Yukawa sector with $10_{\rm H}$ and $126_{\rm H}$ 
Higgs representations is, in terms of fields, the most economical choice that can accommodate all known low-energy data. 
To perform this task, one has either to extrapolate the values of the fermion parameters at the EW scale to the grand unified 
theory (GUT) scale, $M_{\rm GUT}$, or in the opposite direction to impose conditions on the Yukawa matrices defined at $M_{\rm GUT}$ 
and evolve them down to $M_{\rm Z}$.

In this work, we will use the latter approach but, contrary to the procedure usually adopted in the literature, 
we explicitly take into account
% A common feature of these analyses is to consider extrapolated values obtained from the SM renormalization group equations (RGEs), thus completely disregarding 
the presence of intermediate gauge groups, characterized by a mass scale $M_{\rm I}$. 
In fact, besides the evolution of the coupling constants, such contributions are expected to modify the evolution of the fermion 
masses and mixing, introducing relations among the Yukawa couplings at the same scale $M_{\rm I}$. 
We quantify the impact of using such new contributions in the renormalization group equations (RGEs) for 
fermion masses and mixings, 
considering an illustrative and simplified SO(10) model with a breaking chain given by \cite{Deshpande:1992au}:
\begin{eqnarray}
\label{chain}
{\rm SO(10)} &\stackrel{M_{\rm GUT} \, - \, 210_{\rm H}}{\longrightarrow}
&4_{\rm C}\, 2_{\rm L}\, 2_{\rm R}\ \stackrel{M_{\rm I} \, -\, 126_{\rm H}}{\longrightarrow}3_{\rm C}\, 2_{\rm L}\, 1_{Y}\ \stackrel{M_{\rm Z} \, - \, 10_{\rm H}}{\longrightarrow} \ 3_{\rm C}\,1_{\rm Y}\,,
\end{eqnarray}
where the symbols should be self-explanatory. 
In the present model, the intermediate gauge group is the Pati--Salam (PS) group 
${\rm SU(4)}\times {\rm SU(2)}_{\rm L}\times {\rm SU(2)}_{\rm R}$, which was introduced in Ref.~\cite{PhysRevD.10.275}, and in the first step, the 
breaking of SO(10) down to the PS group is achieved by means of a $210_{\rm H}$ representation of Higgs. In the next step, the 
breaking of the PS group down to the SM gauge group is performed by means of a $\overline{126}_{\rm H}$. At $M_{\rm Z}$, the final step 
of the breaking of the SM gauge group to ${\rm SU(3)_{\rm C}}\times {\rm U(1)_{\rm Y}}$ is obtained with a $10_{\rm H}$, we will, 
however, not consider any RG running below $M_{\rm Z}$. Given the exploratory character of our study, we do not address other 
relevant open problems in SO(10) models, such as the presence of a good dark matter candidate in the scalar spectrum or the 
possibility of producing the correct amount of baryon asymmetry in the Universe. We will, however, pay much attention to the energy 
of the GUT scale $M_{\rm GUT}$, the related coupling constant $\alpha_{\rm GUT}$, and the energy of the intermediate scale $M_{\rm I}$, 
since they are all necessary ingredients for a correct evolution of fermion masses and mixings. The output of our analysis will be 
the values of the elements of the Yukawa matrices at $M_{\rm GUT}$, which give a reasonable fit to
the fermion observables at $M_{\rm Z}$. 
These values can directly be compared to the corresponding ones obtained from an evolution without the 
intermediate scale starting 
at $M_{\rm GUT}$, thus allowing a quantification of the new effects introduced by the PS gauge group.

This paper is organized as follows. In Sect.~\ref{notation}, we present our notation for the relevant fields 
and discuss the evolution of the gauge coupling constants. 
Then, in Sect.~\ref{yukawas}, we investigate the renormalization group running of the various Yukawa 
couplings such as the ones for charge leptons, neutrinos, and Higgs self-couplings. 
Next, in Sect.~\ref{procedure}, we present a numerical parameter-fitting procedure to determine the renormalization 
group running of quark and lepton observables from the GUT scale $M_{\rm GUT}$ down to the EW scale $M_{\rm Z}$ and 
to find the effect of the intermediate energy scale $M_{\rm I}$. 
In Sect.~\ref{results}, we give the numerical results and discuss the obtained results. 
Finally, in Sect.~\ref{summary}, we summarize the results and present our conclusions. 
In addition, in Appendix~\ref{app:A}, we list some useful RGEs for our investigation.

\section{Evolution of gauge coupling constants}
\label{notation}

We work in the framework of SO(10) with two representations of Higgs fields, namely the $10_{\rm H}$ and the 
$126_{\rm H}$, which are both relevant for generating the fermion mass matrices. In the PS group, the Higgs and matter 
fields decompose as
\bea
10_{\rm H} &=& (1,2,2) \oplus (6,1,1) \, ,\nn \\
16 &=& (4,2,1) \oplus (\overline{4},1,2) \equiv F_{\rm L} + F_{\rm R}\, , \nn \\
\overline{126}_{\rm H}&=& (6,1,1) \oplus (10,1,3) \oplus (\overline{10},3,1) \oplus (15,2,2) \,,
\eea
where $F_{\rm L}$ and $F_{\rm R}$ are the left- and right-handed parts of the $16$, respectively.
It is useful to introduce the following short-hand notations
\bea
\Phi \equiv (1,2,2)\,, \qquad \Sigma \equiv (15,2,2)\,, \qquad \overline{\Delta_R} \equiv (10,1,3)\,.
\eea
These are the components of $\overline{126}_{\rm H}$ and $10_{\rm H}$ which are involved in the breaking chain. 
It is thus clear that the other components must live at the GUT scale in order not to affect the breaking pattern 
\cite{Altarelli:2013aqa}. For the sake of simplicity, we will restrict ourselves to one-loop matching
so that the evolution equations for the gauge coupling constants $\alpha_i$ between two energy scales $M_1$ and $M_2$ are  
given by the standard formula \cite{Koh:1983ir,PhysRevD.25.581}
\bea
\alpha_i^{-1} (M_2) = \alpha_i^{-1} (M_1) -\frac{a_i}{2\pi} \log \left(\frac{M_2}{M_1}\right) \,,
\eea
where the coefficients $a_i$ can be obtained from, e.g., Ref.~\cite{Koh:1983ir}.
% The coefficients $a_i$ are group theoretical quantities that, for a generic SU($N$), can be written as
% \bea
% a_i = \frac{4}{3} n_g - \frac{11}{3} N + \frac{1}{3}\,\eta \,S_2(R_p)\,,
% \eea
% where $n_g$ is the number of fermion generations, $N$ is the dimension of the group SU($N$), and $\eta=1,1/2$ for complex and real scalar fields, respectively. Given a scalar field $S$, which transforms according to the representation $R=R_1\otimes ... \otimes R_{N'}$, where $R_p$ is an irreducible representation of the group $G_p$ of dimension $d(R_p)$, the factor $S_2(R_p)$ is defined in e.g.~Ref.~\cite{DiLuzio:2011my} as
% \be
% S_2(R_p) \equiv T(R_p)\frac{d(R)}{d(R_p)}\,,
% \label{S2}
% \ee
% where $T(R_p)$ is the Dynkin index of $R_p$ \cite{Koh:1983ir}. 
At $M_{\rm GUT}$, the gauge couplings are unified and the matching conditions are simply
\be
\alpha_{\rm 4C}(M_{\rm GUT})=\alpha_{\rm 2R}(M_{\rm GUT}) =\alpha^\prime_{\rm 2L}(M_{\rm GUT})\,,
\ee
where $\alpha_{\rm 4C}$, $\alpha_{\rm 2R}$, and $\alpha^\prime_{\rm 2L}$ are the coupling constants of SU(4),
${\rm SU(2)}_{\rm R}$, and ${\rm SU(2)}_{\rm L}$ above $M_{\rm I}$, respectively. Next, we have to determine the 
running for $a_{\rm 4C},a^\prime_{\rm 2L}$, and $a_{\rm 2R}$ between $M_{\rm GUT}$ and $M_{\rm I}$. 
The relevant Higgs fields, which are participating in the running in this energy region, are $\Phi$, $\Sigma$, and $\overline{\Delta_R}$. Here, $\Phi$ and $\overline{\Delta_R}$  contribute to $\alpha_{\rm 4C}$, $\Phi$ and $\Sigma$ to $\alpha_{\rm 2L}$, and all three of them to $\alpha_{\rm 2R}$. At $M_{\rm I}$, we impose the relations \cite{Deshpande:1992au}:
\bea
\alpha_{\rm 3C}(M_{\rm I}) &=&\alpha_{\rm 4C}(M_{\rm I})\,, \qquad \alpha_{2L}(M_{\rm I}) =\alpha^\prime_{\rm 2L}(M_{\rm I})\,, 
\qquad \alpha^{-1}_{\rm 1Y}(M_{\rm I}) = \frac{3}{5}\alpha^{-1}_{\rm 2R}(M_{\rm I}) + \frac{2}{5}\alpha^{-1}_{\rm 4C}(M_{\rm I})\nn\,,\\
\label{evolsm}
\eea
where $\alpha_{\rm 3C}$, $\alpha_{\rm 2L}$, and $\alpha_{\rm 1Y}$ are the SM gauge coupling constants. 

Eventually, in the running from $M_{\rm I}$ down to $M_{\rm Z}$, the Higgs representations involved in the RGEs are [under the SM gauge group ${\rm SU(3)}_{\rm C}\times {\rm SU(2)}_{\rm L} \times {\rm U(1)}_{\rm Y}$]
\bea
\Phi &=& (1,2)_{1/2} \oplus (1,2)_{-1/2} = H_u + H_d =\phi_1 + \phi_3\,, \nn \\
\Sigma &=& (1,2)_{1/2} \oplus (1,2)_{-1/2} = H^\prime_u + H^\prime_d =\phi_2 + \phi_4\,.
\eea
%{\color{red}(Should there be $\ni$ or $=$?)}
Hence, in this model, there are four Higgs SU(2) doublets to be dealt with. It is, however, beyond the scope of the present work to discuss in detail the scalar potential of the model and to identify which combination of potential parameters allows a unique light scalar Higgs particle, in accordance with the recent discovery at the LHC.

At $M_{\rm Z} \simeq 91.19$~GeV, imposing the experimental constraints, given by \cite{Amsler:2008zzb}
\bea
\alpha_{3C}(M_{\rm Z}) &=&  0.1176 \pm 0.002\,, \nn \\ 
\alpha_{2L}(M_{\rm Z}) &=& 0.033812 \pm 0.000021\,, \nn \\
\alpha_{1Y}(M_{\rm Z})&=& 0.016946 \pm 0.000006\,,
\eea
we obtain the values of the mass scales (to be used throughout this work):
\bea 
M_{\rm I} = (1.5 \pm 0.2)\cdot 10^{12} \,\text{GeV}\,, \qquad 
M_{\rm GUT} = (1.7 \pm 0.6)\cdot 10^{15} \,\text{GeV}\,,
\eea
and the value of the gauge coupling at $M_{\rm GUT}$ is $\alpha_{\rm GUT} \simeq 0.027$.

The errors on the mass scales only include the propagated uncertainties from the SM coupling constants and the Z boson mass. 
Then, although the value of $M_{\rm GUT}$ is marginally compatible with a naive estimate of the life-time of the proton, 
which would require $M_{\rm GUT} \sim 10^{16}$~GeV, unknown threshold corrections \cite{Mohapatra:1992dx} can easily increase 
the estimated errors, thus 
allowing for a larger value of $M_{\rm GUT}$. We should stress again that the main goal of this work is to quantify the 
effects of $M_I$ on the RGEs for the fermion observables rather than to construct a realistic model based on SO(10).

\section{RG running of Yukawa couplings}
\label{yukawas}

In this section, we briefly discuss the relevant RGEs and matching conditions among the Yukawa couplings defined at the 
SO(10) breaking scale, {\it i.e.}~the GUT scale, and at the intermediate scale.

\subsection{Charged leptons}

At $M_{\rm GUT}$, the Yukawa sector reads
\bea
\label{yuk}
L_Y=16\,(h\,10_{\rm H}+f\,\overline{126}_{\rm H})\,16\,,
\eea
where $h$ and $f$ are unknown symmetric couplings to be determined through a fitting procedure. 
Furthermore, in the region between $M_{\rm GUT}$ and $M_{\rm I}$, the Yukawa part of the Lagrangian is given by \cite{Fukuyama:2002vv}
\begin{eqnarray}
 -{\mathcal L}_{\rm Y} &=&\sum_{i,j}\left( Y^{(10)}_{ F\,ij}F_{\rm L}^{iT}\Phi F_{\rm R}^{j}
  +Y^{(126)}_{F\,ij}F_{\rm L}^{iT}\Sigma F_{\rm R}^{j}
+Y^{(126)}_{R\,ij}F_{\rm R}^{iT}\overline{\Delta_R} F_{\rm R}^{j}+\text{h.c.}\right)\,,
\label{y2}
\end{eqnarray}
where $Y^{(10)}_F$, $Y^{(126)}_F$, and $Y^{(126)}_R$ are Yukawa couplings.
In this region, the one-loop RGEs for the effective Yukawa couplings  have been computed in Ref.~\cite{Fukuyama:2002vv} and 
given for reference in Appendix~\ref{app:A1}. Furthermore, at $M_{\rm GUT}$, the couplings $Y^{(10)}_F$, $Y^{(126)}_F$, 
and $Y^{(126)}_R$ have to be matched to $h$ and $f$:
\bea
&& \frac{1}{\sqrt{2}}Y^{(10)}_F (M_{\rm GUT}) \equiv h\,, \nn \\
&& \frac{1}{4\sqrt{2}}Y^{(126)}_F(M_{\rm GUT}) =\frac{1}{4}Y^{(126)}_R (M_{\rm GUT})
\equiv f\,, \label{bc1}
\eea
where the numerical factors are Clebsch--Gordan coefficients needed for a correct embedding of PS into SO(10) \cite{Aulakh:2002zr}. Note that in order to derive the fermion mass matrices one has to introduce the vev's of the appropriate Higgs multiplets. In standard notation, the relevant contribution to fermion masses and mixing come from the $\Phi$ submultiplet of the $10_{\rm H}$ and the $\Sigma$ submultiplet of the $126_{\rm H}$, which can be written as
\be
k_{u,d} \equiv \langle \Phi_{u,d} \rangle_{10}\,, \qquad v_{u,d} \equiv \langle \Sigma_{u,d} \rangle_{126}\,.
\label{kvvevs}
\ee
In particular, it is useful to introduce the ratios
\bea
r_v \equiv \frac{k_u}{k_d}\,, \qquad s \equiv \frac{v_u}{r_v v_d}\,, \label{ratios}
\eea
which allow us, using the Lagrangian given in Eq.~\eqref{yuk} and the previous definitions, to obtain the following fermion mass matrices \cite{Dutta:2004zh,Dutta:2005ni,Altarelli:2010at,Joshipura:2011nn,Dueck:2013gca}
\bea\label{masses}
&& M_u = h\,k_u + f\,v_u\,, \qquad M_d = h\,k_d + f\,v_d\,, \nn \\
&& M_D = h\,k_u -3\, f\,v_u\,, \qquad M_e = h\,k_d -3\, f\,v_d\,, \qquad M_R = f\,v_R\,,
\eea
where $M_u$, $M_d$, $M_D$, $M_e$, and $M_R$ are the up-type quark, down-type quark, Dirac, charged-lepton, and right-handed neutrino mass matrices, respectively, and $v_R=\left\langle \overline{\Delta_R} \right\rangle$ is the vev of $\overline{\Delta_R}$. Using the relations in Eqs.~\eqref{bc1}--\eqref{ratios}, we can rewrite Eq.~\eqref{masses} as
\bea\label{massnew}
M_u &=& \frac{r_v}{\sqrt{2}} \, \left(k_d Y_F^{(10)} + \frac{v_d \,s}{4} Y_F^{(126)} \right)\,, \nn \\
M_d &=& \frac{k_d}{\sqrt{2}} Y_F^{(10)} + \frac{v_d}{4\sqrt{2}} Y_F^{(126)}\,, \nn \\
M_e &=& \frac{k_d}{\sqrt{2}} Y_F^{(10)} -3 \frac{v_d}{4\sqrt{2}} Y_F^{(126)}\,, \nn \\
M_D &=& \frac{r_v}{\sqrt{2}} \, \left(k_d Y_F^{(10)} -3  \frac{v_d \,s}{4} Y_F^{(126)} \right)\,.
\eea
Finally, in the region between $M_{\rm I}$ and $M_{\rm Z}$, the RG running down to the SM produces formally equivalent mass matrices, where we only have to distinguish among the upper and lower component of the ${\rm SU(2)}_{\rm L}$ doublets:
\bea\label{massfin}
M_u &=& \frac{r_v}{\sqrt{2}} \, \left(k_d Y_u^{(10)} + \frac{v_d \,s}{4} Y_u^{(126)} \right)\,, \nn \\
M_d &=& \frac{k_d}{\sqrt{2}} Y_d^{(10)} + \frac{v_d}{4\sqrt{2}} Y_d^{(126)}\,, \nn \\
M_e &=& \frac{k_d}{\sqrt{2}} Y_e^{(10)} -3 \frac{v_d}{4\sqrt{2}} Y_e^{(126)}\,, \nn \\
M_D &=& \frac{r_v}{\sqrt{2}} \, \left(k_d Y_\nu^{(10)} -3  \frac{v_d \,s}{4} Y_\nu^{(126)} \right)\,,
\eea
with the matching conditions at $M_{\rm I}$ given by
\begin{eqnarray}\label{matchmi}
&& Y^{(10)}_u (M_{\rm I})=Y^{(10)}_d(M_{\rm I})=Y^{(10)}_{\nu}(M_{\rm I})
=Y^{(10)}_e(M_{\rm I})\equiv Y^{(10)}_F\,, \nn \\
&& Y^{(126)}_u(M_{\rm I})=Y^{(126)}_d(M_{\rm I})=-\frac{1}{3}Y^{(126)}_{\nu}(M_{\rm I})
=-\frac{1}{3}Y^{(126)}_e(M_{\rm I})\equiv Y^{(126)}_F\,,
\end{eqnarray}
where the factor $1/3$ is a consequence of the property of the vev $\langle \Sigma \rangle \sim {\rm diag} (1,1,1,-3)$. In this region, the Yukawa Lagrangian is given by
\begin{eqnarray}
-{\mathcal L}_{Y}&=&\sum_{i,j}\left(Y^{(10)}_{u\,ij}\,
\overline{q^i_{\rm L}}\,\widetilde{\phi_1}u_{\rm R}^j
+Y^{(126)}_{u\,ij}\,\overline{q^i_{\rm L}}\,\widetilde{\phi_2}u_{\rm R}^j 
+Y^{(10)}_{d\,ij}\,\overline{q^i_{\rm L}}\,\phi_3d_{\rm R}^j
+Y^{(126)}_{d\,ij}\,\overline{q^i_{\rm L}}\,\phi_4d_{\rm R}^j \right. \nn\\
 &+&\left. Y^{(10)}_{\nu\, ij}\,\overline{\ell^i_{\rm L}}\,\widetilde{\phi_1}N_{\rm R}^j
+Y^{(126)}_{\nu\, ij}\,\overline{\ell^i_L}\,\widetilde{\phi_2}N_R^j
+Y^{(10)}_{e\,ij}\,\overline{\ell^i_{\rm L}}\,\phi_3e_{\rm R}^j
+Y^{(126)}_{e\,ij}\,\overline{\ell^i_{\rm L}}\,\phi_4e_{\rm R}^j+\text{h.c.}
\right)\,,
\label{y}
\end{eqnarray}
where $q_{\rm L}$ and $\ell_{\rm L}$ are the usual quark and lepton ${\rm SU(2)}$ doublets, respectively, and $u_{\rm R}$, $d_{\rm R}$, and $e_{\rm R}$ are the corresponding ${\rm SU(2)}$ singlets, and $N_{\rm R}$ is the right-handed neutrino field. The RGEs of the Yukawa couplings from $M_{\rm I}$ to $M_{\rm Z}$ are presented in Appendix~\ref{app:A2}. To all Higgs fields, one can assign vev's $\phi_i = v_i/\sqrt{2}$, which, 
in terms of the vev's of Eq.~\eqref{kvvevs}, read
\bea
\label{vek}
v_1 = k_u\,, \qquad 4 v_2 = v_u\,, \qquad v_3 = k_d\,, \qquad 4 v_4 = v_d \,.
\eea
In our investigation, for the sake of simplicity, these vev's will be considered as fixed quantities.

\subsection{Neutrinos}
\label{sec:neu}

In the RG running of the Yukawa couplings, a further complication arises from the fact that besides the 
intermediate energy scale $M_{\rm I}$, there are also three seesaw energy scales related to the three 
heavy right-handed neutrinos which need to be taken into account. The picture can be simplified by assuming that all 
heavy neutrinos obtain the same mass at a seesaw energy scale coinciding with $M_{\rm I}$. 
% It does in fact turn out to be a 
% rather good choice, which has been numerically verified in Ref.~\cite{Altarelli:2013aqa}. 
In order to define the concept of neutrino masses and leptonic mixing as functions of the renormalization scale $\mu$, we use the standard see-saw formula:
\bea
m_\nu(\mu) = M_D^T(\mu)\, M_R^{-1}(\mu)\, M_D(\mu)\,,
\label{full}
\eea
where $M_D$ and $M_R$ are $\mu$-dependent quantities.
% and $v=256$ GeV {\color{green}assumed, as the other vev's, independent 
% on the energy scale} .
% The {\color{green} meaning} of the quantities $y_\nu(\mu)$ and  $M_R(\mu)$  depends on the energy range under discussion. 
In particular, above the seesaw energy scale, {\it i.e.}~above the intermediate scale where $\mu>M_{\rm I}$, 
the matrix $M_R(\mu)$ in Eq.~\eqref{full} is a RG running quantity defined as
\begin{equation} 
M_R=\frac{1}{4}\left\langle \overline{\Delta_R} 
\right\rangle Y^{(126)}_R\,.
\label{triplet}
\end{equation}
Hence, assuming that $\langle \overline{\Delta_R}\rangle$ is a $\mu$-independent quantity,
its evolution is fully determined by the evolution of $Y^{(126)}_R$. In this energy region, $M_D(\mu)$ is given by the Dirac mass matrix in Eq.~\eqref{massnew}, and thus, 
it obtains contributions from 
$Y_F^{(10)}$ and $Y_F^{(126)}$. Inserting the expressions for $M_D$ and $M_R$ into the seesaw relation in Eq.~\eqref{full} for the light neutrino masses, we obtain
\bea\label{neutrinomass}
m_{\nu} &=& \frac{r_v^2}{2}\left(k_d^2 Y_F^{^T(10)} M_R^{-1} Y_F^{(10)}- 3 \frac{v_d k_d s }{4}Y_F^{^T(126)} M_R^{-1} Y_F^{(10)}
\right.\nn \\ &-& \left. 3 \frac{v_d k_d s }{4}Y_F^{^T(10)} M_R^{-1} Y_F^{(126)} + \frac{9 v_d^2 s^2}{16}Y_F^{^T(126)} M_R^{-1} Y_F^{(126)}
\right)\,.
\eea

Below the seesaw scale, {\it i.e.}~below the intermediate scale where $\mu < M_{\rm I}$, we will instead consider 
an effective neutrino mass operator:
\begin{eqnarray}\label{fourpoint}
{\mathcal L}_{\nu}&=&\frac{1}{4}\sum_{a,b=1,2}
\sum_{i,j}\kappa^{(a,b)}_{ij}\left(\overline{\ell^i_{\mathrm{L}\delta}}
\epsilon_{\delta\gamma}\widetilde{\phi_{a\mathrm{\gamma}}}\right)\left(\phi^{*}_{b\alpha}
\epsilon^{\alpha\beta}\ell^{Cj}_{{\rm L}\beta}\right)+\text{h.c.}\,,
\end{eqnarray}
where $\kappa^{(a,b)}_{ij}$ are flavor matrices satisfying 
$\kappa^{(a,b)}_{ij}=\kappa^{(b,a)}_{ji}$ \cite{Grimus:2004yh}
% is the effective neutrino mass matrix 
and $\epsilon_{\alpha\beta}$ is the two-dimensional antisymmetric tensor.
Now, the light neutrino mass matrix can be expressed in terms of these effective coefficients
% , which are elements of the quantity denoted $\kappa$, 
and it is thus given by
\begin{equation}\label{Mnu}
m_{\nu}=\frac{1}{2}\sum_{a,b=1,2}\kappa^{(a,b)}v_a^{*}v_b^{*}\,.
\end{equation}
Then, we can construct the matching conditions at $M_{\rm I}$. 
This is performed by comparing Eq.~\eqref{Mnu} with Eq.~\eqref{neutrinomass} using the relations in Eq.~\eqref{vek}, which then gives the following expressions
\bea\label{kappas}
\kappa^{(1,1)} &=& Y_F^{^T(10)} M_{\rm R}^{-1} Y_F^{(10)}\,, \nn \\
\kappa^{(1,2)}&=& -3 Y_F^{^T(126)} M_{\rm R}^{-1} Y_F^{(10)}\,, \nn \\
\kappa^{(2,1)}&=& -3 Y_F^{^T(10)} M_{\rm R}^{-1} Y_F^{(126)}\,, \nn \\
\kappa^{(2,2)}&=& 9 Y_F^{^T(126)} M_{\rm R}^{-1} Y_F^{(126)} \,.
\eea
The RGEs for the coefficients $\kappa^{(a,b)}_{ij}$ are 
% given in Ref.~\cite{Fukuyama:2002vv} and 
presented for reference in Appendix~\ref{app:A3}. These RGEs depend on the Higgs self-couplings $\lambda_{ijkl}$, 
which need to be taken into consideration in a consistent way. We will discuss these Higgs self-couplings in the next subsection.

\subsection{Higgs self-couplings}

As previously observed, our model contains four Higgs doublets at low energies, 
two doublets from $\Phi$, {\it i.e.}~$\phi_1$ and $\phi_3$, and two from $\Sigma$, {\it i.e.}~$\phi_2$ and $\phi_4$. 
The doublets $\phi_1$ and $\phi_2$ couple to up-type quarks and leptons, whereas the doublets $\phi_3$ and 
$\phi_4$ couple to down-type quarks and leptons. 
The quartic terms in the scalar potential have the general form
\bea
V\left(\phi \right)&=&\frac{1}{4!}\sum_{a,b,c,d=1,2,3,4}\lambda_{abcd} 
\left(\phi_a^{\dagger} \phi_b \right)
\left(\phi_c^{\dagger} \phi_d \right)\,.
\label{higgspotential}
\eea
This is a rather tedious expression which can, however, be simplified using the fact that the quartic couplings obey the 
following relations [derived from the form of the potential in Eq.~\eqref{higgspotential}]
\be
\lambda_{abcd}=\lambda_{cdab}=\lambda_{badc}^*\,.
\ee
The RGEs for the Higgs self-couplings are given by \cite{Cheng:1973nv}
\bea
16\pi^2\frac{\dr \lambda_{abcd}}{\dr t}&=&\frac{1}{6}\sum_{m,n=1,2,3,4}\left(2\lambda_{abmn}\lambda_{nmcd}+\lambda_{abmn}\lambda_{cmnd}+\lambda_{amnb}\lambda_{mncd}\right.\nonumber\\ 
&+& \left. \lambda_{amnd}\lambda_{cnmb}+\lambda_{amcn}\lambda_{mbnd}\right) -3(3g_2^2+g_Y^2)\lambda_{abcd}\nonumber\\
&+& 9 (3g_2^4+g_Y^4)\delta_{ab}\delta_{cd}+36g_2^2g_Y^2\left(\delta_{ad}\delta_{bc}-\frac{1}{2}\delta_{ab}\delta_{cd}\right)\nonumber\\
&+& \sum_{m,n=1,2,3,4}(\lambda_{mbcd}A_{am}+\lambda_{amcd}A_{mb}+\lambda_{abmd}A_{cm}+\lambda_{abcm}A_{md})\nonumber\\
&-& 48H_{abcd},\hspace{2mm}(a,b,c,d=1,2,3,4)\,,
\eea 
where we have defined the auxiliary quantities
\bea
A_{ab}\equiv \mathrm{tr}(3Y_a^{u\dagger}Y_B^{u\phantom\dagger}+3Y_a^{d\dagger}Y_b^{d\phantom\dagger}+Y_a^{e\dagger}Y_b^{e\phantom\dagger})
\eea
and
\bea
H_{abcd}&\equiv & \mathrm{tr}(3Y_d^{u\dagger}Y_c^{u\phantom\dagger}Y_b^{u\dagger}Y_a^{u\phantom\dagger}+3Y_a^{d\dagger}Y_b^dY_c^{d\dagger}Y_d^{d\phantom\dagger}
+Y_a^{e\dagger}Y_b^{e\phantom\dagger}Y_c^{e\dagger}Y_d^{e\phantom\dagger}\nonumber\\
&+& 3Y_a^{u\dagger}Y_b^{u}Y_d^{d\dagger}Y_c^{d}+3Y_b^{d\dagger}Y_a^{d}Y_c^{u\dagger}Y_d^{u}-3Y_d^{d\dagger}Y_c^{d}Y_b^{u\dagger}Y_a^{u}
-3Y_a^{u\dagger}Y_d^{u}Y_b^{d\dagger}Y_c^{d})\,.
\eea
Furthermore, the following abbreviations have been used
\bea
&& Y_1^{u} \equiv Y_u^{(10)}\,, \qquad Y_2^{u} \equiv Y_u^{(126)}\,, \qquad Y_3^d \equiv Y_d^{(10)}\,, \qquad Y_4^d \equiv Y_d^{(126)}\,,\nn \\
&& Y_3^{e} \equiv Y_e^{(10)}\,, \qquad Y_4^{e} \equiv Y_e^{(126)}\,, \qquad \text{otherwise zero}\,.
\eea
Above $M_{\rm I}$, there are four distinct Higgs self-couplings $\lambda_i$ ($i = 1,2,3,4$), 
which are matched to the low-energy counterparts at $M_{\rm I}$ as
\bea
\frac{\lambda_{abcd}}{4!}&=&\lambda_1 \hspace{2mm}\text{for }a,b,c,d=\{1,3\}\,,\nn \\
\frac{\lambda_{abcd}}{4!}&=&2\lambda_2 \hspace{2mm}\text{for } a,b=\{1,3\} \text{ and } c,d=\{2,4\}\,, \nn \\
\frac{\lambda_{abcd}}{4!}&=&\lambda_3\hspace{2mm}\text{for }\{a,b\},\{ c,d\}=\{1,2\}, \{1,4\}, \{3,2\},\{3,4\}\,, \nn \\
\frac{\lambda_{abcd}}{4!}&=&4\lambda_4 \hspace{2mm}\text{for }=\{2,4\}\,.
\eea
Note that the RG running of the Higgs couplings above $M_{\rm I}$ is irrelevant for the evolution of the fermion masses and mixing parameters and therefore not taken 
into account here. 
In order to perform a numerical computation of the RG running of 
the fermionic parameters, one has to specify the choice of the initial conditions for the Higgs couplings. 
For the sake of simplicity, we allowed one of the Higgs couplings, $\lambda_1$, to be free and the other three 
were fixed to $\lambda_2 = 2\cdot 10^{-2}$, $\lambda_3 = 1\cdot 10^{-4}$, and $\lambda_4 = 4\cdot 10^{-3}$.

\section{Numerical parameter-fitting procedure}
\label{procedure}

In this section, we present the numerical strategy that we have used to show the effect of the intermediate scale $M_{\rm I}$ on 
the extrapolated values of fermion masses and mixings from the GUT scale $M_{\rm GUT}$. As we previously explained, we adopt 
the procedure of considering the entries of the couplings $h$ and $f$ as well as the vevs as our free parameters and 
evolving them down to the EW scale $M_{\rm Z}$, where the values of masses and mixings of quarks, charged leptons, and neutrinos 
are known. There are in total 19 free parameters at $M_{\rm GUT}$ which need to be determined,  including one Higgs 
self-coupling at $M_{\rm I}$. Without loss of generality, 
we can work in the basis where the Yukawa coupling matrix $h$ is real and diagonal. 
Then, we have three parameters in the real diagonal Yukawa coupling matrix $h$, twelve in the complex and symmetric Yukawa coupling matrix $f$, one in the parameter $r_v$ (which can be chosen to be real), two in the complex parameter $s$, and finally one in the vevs $k_d=v_d$. 
In addition we shall fit  one of the Higgs couplings $\lambda_1$ at the intermediate scale.

The evolved observables depend on all the parameters, so an analytical minimization of the $\chi^2$ function is not feasible. Hence, we adopt a numerical strategy, which consists of the following steps:
\begin{itemize}
\item First, the values of the parameters at $M_{\rm GUT}$ are randomly generated according to some prior distribution.
\item Then, they are evolved down to $M_{\rm Z}$ after solving the RGEs discussed in previous sections.
\item Next, at $M_{\rm Z}$, the observables can then be constructed and compared to experimental data.
\item Finally, the procedure is repeated with new randomly sampled parameter values from a reduced parameter space and the result 
is given when convergence on the point with largest likelihood occurs, {\it i.e.}~the best-fit point.
\end{itemize}
The advantage of using such a sampling algorithm rather than a simple parameter scan is that it is significantly more computationally efficient. For the sampling procedure, we used the software MultiNest, which is based on nested sampling normally used for calculation of the Bayesian evidence \cite{Feroz:2007kg,Feroz:2008xx,Feroz:2013hea}. Nested sampling reduces the many-dimensional integration of the likelihood to a one-dimensional integral, which significantly will increase the speed of the calculation \cite{Skilling:2004,Skilling:2006}. 
The sampling space is reduced for each iteration, removing points with small values of the likelihood. 
Thus, in each step of iteration, we will replace the points with the smallest values of the likelihood by 
points with larger values of the likelihood. Eventually, we will find the point with the largest value of the 
likelihood, which is then the point that we will use for the fit. This point is called the best-fit point. Since this method is Bayesian, we necessarily have to make a choice of prior distributions for the parameters, which are fitted at $M_{\rm GUT}$. Note that we are not interested in the Bayesian analysis as such, and therefore, these priors could be considered simply as a bound on the parameter space. Nevertheless, since the orders of magnitude were unknown for the parameters in the matrices $h$ and $f$, it was relevant to use logarithmic priors, ranging from $10^{-15}$ to $10^{-1}$. For the remaining parameters suitable uniform priors were used. The comparison to the EW data is performed by maximizing the value of the logarithm of the likelihood $\mathcal{L}$, which to a rather good approximation, {\it i.e.}~the Gaussian approximation, 
is related to the $\chi^2$ through
% \be
$\chi^2 = -2\log (\mathcal{L})$.
% \ee
The $\chi^2$ function is, as usual, defined as
\be
\chi^2 \equiv \sum_{i=1}^n \left(\frac{X_i-\mu_i}{\sigma^{\rm exp}_i}\right)^2\,,
\label{eq:chi2}
\ee
where $X_i$ is the experimental value of the $i$th observable, $\mu_i$ the expectation value from the model, and $\sigma_i^{\rm exp}$ the experimental uncertainty. All observables which were used are presented in Table~\ref{tab:quantities}. 
\begin{table}[ht!]
\centering
\begin{tabular}{| l | c | c | l | c | c |}
\hline\hline
\multicolumn{3}{|c|}{Quark sector } & \multicolumn{3}{|c|}{Lepton sector}\\
\hline
Observable&$X_i$&$\sigma^{\rm exp}_i$&Observable&$X_i$&$\sigma^{\rm exp}_i$\\
\hline
$m_{d}$ (GeV)& $2.9\cdot 10^{-3}$&$1.215\cdot 10^{-3}$&$m_{e}$ (GeV)&$4.8657\cdot 10^{-4}$ &$2.4339\cdot 10^{-5}$ \\
$m_{s}$ (GeV)& $5.5\cdot 10^{-2}$& $1.55\cdot 10^{-2}$&$m_{\mu}$ (GeV)&$1.0272\cdot 10^{-1}$&$5.14\cdot 10^{-3}$\\
$m_{b}$ (GeV) &2.89& $9.0\cdot 10^{-2}$&$m_{\tau}$ (GeV)&$1.74624$ &$8.731\cdot 10^{-2}$\\
$m_{u}$ (GeV)&$1.27\cdot 10^{-3}$&$4.6\cdot 10^{-4}$&$r\equiv\frac{\Delta m_{21}^2}{\Delta m_{31}^2}$ &$0.030$&$0.0033$\\
$m_{c}$ (GeV)&$6.19\cdot 10^{-1}$&$8.4\cdot 10^{-2}$&$\sin^2\theta^\ell_{12}$&0.30&$1.3\cdot 10^{-2}$\\
$m_{t}$ (GeV)&$171.7$&$3.0$&$\sin^2\theta^\ell_{13}$&$2.3 \cdot 10^{-2}$&$2.3 \cdot 10^{-3}$\\
$\sin\theta^{q}_{12}$&$2.246\cdot 10^{-1}$&$1.1\cdot 10^{-3}$&$\sin^2\theta^\ell_{23}$&0.41&$3.1 \cdot 10^{-2}$\\
$\sin\theta^{q}_{13}$&$3.5\cdot 10^{-3}$&$3\cdot 10^{-4}$&&&\\
$\sin\theta^{q}_{23}$&$4.2\cdot 10^{-2}$&$1.3\cdot 10^{-3}$&&&\\
$\delta_{\rm CKM}$& $1.2153$ & $5.76\cdot 10^{-3}$&&&\\
\hline
\hline
\end{tabular}
\caption{\it The observables used in the $\chi^2$ for parameter fit at the GUT scale. The experimental values $\{X_i\}$ of the observables are the values of the observables at the EW scale and the values $\{\sigma^{\rm exp}_i\}$ are the respective experimental uncertainties. The values of the quark and charged lepton masses are taken from Ref.~\cite{Xing:2007fb}, the quark mixing parameters from Ref.~\cite{Joshipura:2011nn}, and the neutrino mass-squared differences and the leptonic mixing angles from Ref.~\cite{GonzalezGarcia:2012sz}.} \label{tab:quantities}
\end{table}

For the purpose of this work, we only consider normal hierarchy (NH) for the neutrino masses.\footnote{This is motivated by the difficulty to perform a proper fit for the inverse hierarchy (IH) for the neutrino masses using models similar to ours\cite{Altarelli:2013aqa,Dueck:2013gca}.} 
In addition, we have not used the experimental uncertainties for the charged leptons, since these errors are very small. The minimization procedure of 
the $\chi^2$ would not converge in a reasonable time using the true experimental errors, 
since even a relatively small deviation from the experimental value would have a large impact on the magnitude of the $\chi^2$. 
In the present investigation, we are not interested in determining the values of the charged lepton masses to a 
great precision but rather to obtain values which are relatively close to the values measured at $M_{\rm Z}$, 
with a precision comparable to that of the measurements on the other SM observables. 
Therefore, we choose to impose a relative error on the charged lepton masses of 5 \%.

Thus, the final result of this procedure will be the determination of the unknown para\-meters and, correspondingly, 
the values of the fermion observables at $M_{\rm GUT}$. The effect of $M_{\rm I}$ on the RG running is appreciated by 
comparing such values with the ones obtained from RG running without $M_{\rm I}$, however still taking the seesaw scale into account.

\section{Numerical results and discussion}
\label{results}

Using the numerical parameter-fitting procedure described in Sect.~\ref{procedure}, we perform a fit of the SO(10) 
model parameters at the GUT scale $M_{\rm GUT}$ such that the experimentally known values of the physical 
fermion observables at the EW scale $M_{\rm Z}$ are reproduced. 
Applying this procedure, the obtained values of the Yukawa couplings at $M_{\rm GUT}$ are:
\bea
h&\simeq&{ \left( {\small \begin{array}{ccc}
5.03\cdot 10^{-5} & 0 & 0 \\
0 & -4.92\cdot 10^{-3} & 0 \\
0 & 0 & 5.54\cdot 10^{-1}\end{array}}  \right)}\,,\label{eq:h}\\
f&\simeq&{ \left(  {\small \begin{array}{ccc}
3.14\cdot 10^{-5}{\rm i}& -7.21\cdot 10^{-4}-5.37{\rm i}\cdot 10^{-5}{\rm i} & -1.31\cdot 10^{-3}\\
-7.21\cdot 10^{-4}-5.37\cdot 10^{-5}{\rm i} & 1.09\cdot 10^{-3}-7.26\cdot 10^{-3}{\rm i}& -6.91\cdot 10^{-5}+2.39\cdot 10^{-2}{\rm i} \\
-1.31\cdot 10^{-3}& -6.91\cdot 10^{-5}+2.39\cdot 10^{-2}{\rm i}& 5.56\cdot 10^{-2}+4.53\cdot 10^{-2}{\rm i}\end{array}}  \right)}\,. \nn \label{eq:f}\\
\eea
The fit of the vevs $k_d$ and $v_d$ was done under the simplifying assumption that $k_d=v_d$ and the best-fit value of this parameter was found to be $k_d=v_d \simeq 
 3.75$~GeV. Furthermore, the best-fit values of the parameters $s$ and $r_v$ were found to be $s \simeq 3.57\cdot 10^{-2}+0.40{\rm i}$ and $r_v \simeq 65.3$, 
 respectively, which means that the best-fit value of the parameter $k_u$ is given by $k_u \simeq 245$~GeV using 
 Eq.~\eqref{ratios}. Since we have the freedom of rescaling the values of the vevs by dividing them with a common 
 factor and multiplying the Yukawa couplings with the same common factor, the fit has been performed in such a 
 way that the sum of the squares of the Higgs field vevs in Eq.~\eqref{vek} is equal to 246 GeV. At the intermediate scale $M_{\rm I}$, the values of the Higgs self-couplings had to be determined. These values are, in principle, arbitrary as long as the correct results are reproduced and the values are below the perturbative limit. Hence, only one of the Higgs self-couplings $\lambda_1$ was part of the fit and was fitted to a value of $\lambda_1 \simeq 8.23\cdot 10^{-4}$, while the rest were kept fixed. The fit resulted in a value of the $\chi^2$ function given in Eq.~\eqref{eq:chi2} 
 that is $\chi^2 \simeq 12.7$, which is reasonable for this fit taking its complexity into account.

The values of the observables in the SO(10) model at the EW scale are given in Table~\ref{tab:EW} together with the corresponding pulls for these observables. In general, the pull is defined as
\begin{equation}
g_i \equiv \frac{X_i-\mu_i}{\sigma^{\rm exp}_i} \,,
\end{equation}
where $X_i$ is the value of the $i$th observable at the EW scale (given in Table~\ref{tab:quantities} for all observables used), $\mu_i$ is the value in the SO(10) model, and $\sigma^{\rm exp}_i$ is the experimental uncertainty (again given in Table~\ref{tab:quantities} for all observables used). The observables which are most difficult to accommodate are the down, strange, and top quark masses as well as the quantities $r$ and $\sin^2\theta^\ell_{12}$, although the experimental values are reproduced within about 2$\sigma$. The absolute neutrino mass scale can be inferred once the vev $v_R$ in Eq.~\eqref{masses} is determined by demanding that the small neutrino mass-squared difference $\Delta m^2_{21}$ (or the large one $\Delta m^2_{31}$) resulting from the fit procedure reproduces the experimental value of $\Delta m^2_{21} = 7.5 \cdot 10^{-5} \, {\rm eV}^2$ (or $\Delta m^2_{31} =  2.5 \cdot 10^{-3} \, {\rm eV}^2$). It turns out that $v_R \simeq 1.3\cdot 10^{14}$~GeV and therefore the neutrino masses will have the following values: $m_{\nu_1} \simeq 8.0\cdot 10^{-3}$~eV, $m_{\nu_2} \simeq 1.2\cdot 10^{-2}$~eV, and $m_{\nu_3} \simeq 4.6\cdot 10^{-2}$~eV. In addition, the values of the leptonic Dirac and Majorana CP-violating phases $\delta$, $\rho$, and $\sigma$ can be predicted (which are independent of $v_R$), they are $\delta \simeq 1.67 \simeq 0.53 \pi$, $\rho \simeq 4.00$, and $\sigma \simeq 3.76$. However, note that the values of these phases are dependent on the best-fit point, and there are several points with similar values of the $\chi^2$, which would give rather different values for the three phases. Nevertheless, the value obtained for leptonic Dirac CP-violating phase $\delta$ can be compared with values of $\delta$ from global fits, which all favor a value of $\delta = 3\pi/2$ \cite{Capozzi:2013csa,Forero:2014bxa,Gonzalez-Garcia:2014bfa}.
\begin{table}[ht!]
\centering
\begin{tabular}{| l | c | c | l | c | c |}
\hline\hline
\multicolumn{3}{|c|}{Quark sector } & \multicolumn{3}{|c|}{Lepton sector}\\
\hline
Observable&$ \mu_i $&$g_i$&Observable&$ \mu_i$&$g_i$\\
\hline
$m_{d}$ (GeV) & $3.6 \cdot 10^{-4}$ & $2.1$ &$m_{e}$ (GeV)&$4.8\cdot 10^{-4}$ &$0.22$ \\
$m_{s}$ (GeV) & $0.037$& $1.1$&$m_{\mu}$ (GeV)&$0.10$&$-0.055$\\
$m_{b}$ (GeV) & 2.9& $0.11$&$m_{\tau}$ (GeV)&$1.7$ &$0.52$\\
$m_{u}$ (GeV) & $1.4\cdot 10^{-3}$&$-0.28$&$r\equiv\frac{\Delta m_{21}^2}{\Delta m_{31}^2}$ &$0.036$&$-1.5$\\
$m_{c}$ (GeV)&$0.68$&$-0.73$&$\sin^2\theta^\ell_{12}$&0.28&$1.5$\\
$m_{t}$ (GeV)&$170$&$1.1$&$\sin^2\theta^\ell_{13}$&$0.022$&$0.41$\\
$\sin\theta^{q}_{12}$&$0.23$&$-0.45$&$\sin^2\theta^\ell_{23}$&0.42&$-0.41$\\
$\sin\theta^{q}_{13}$&$0.0035$&$0.0$&&&\\
$\sin\theta^{q}_{23}$&$0.042$&$0.078$&&&\\
$\delta_{\rm CKM}$& $1.2$ & $-0.029$&&&\\
\hline
\hline
\end{tabular}
\caption{\it The values $\{\mu_i\}$ of the observables in the ${\rm SO(10)}$ model at the EW scale presented together with the respective pulls $\{g_i\}$.} \label{tab:EW}
\end{table}

\begin{figure}
\begin{center}
\subfigure{\includegraphics[scale = 0.38]{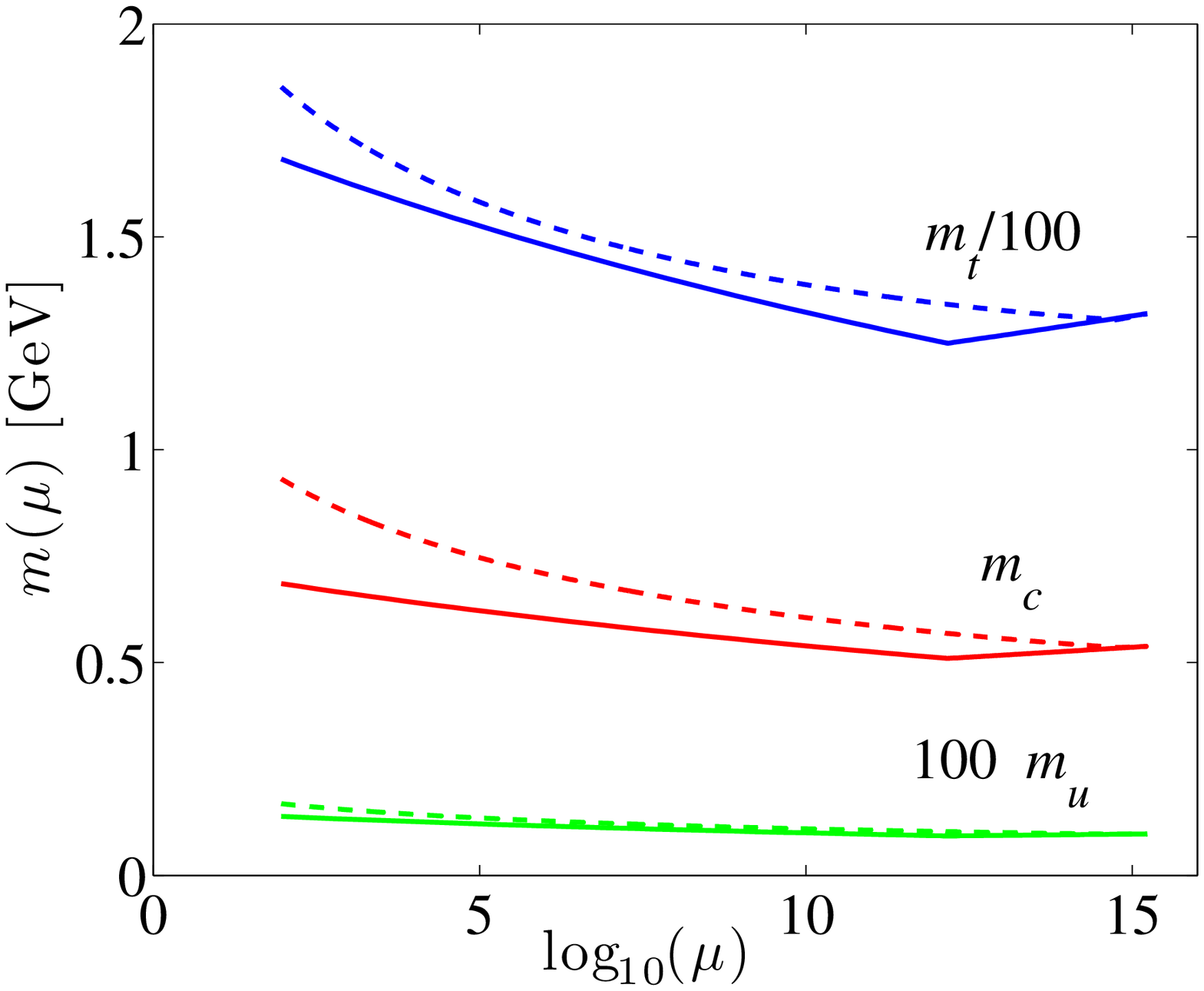}}
\subfigure{\includegraphics[scale = 0.38]{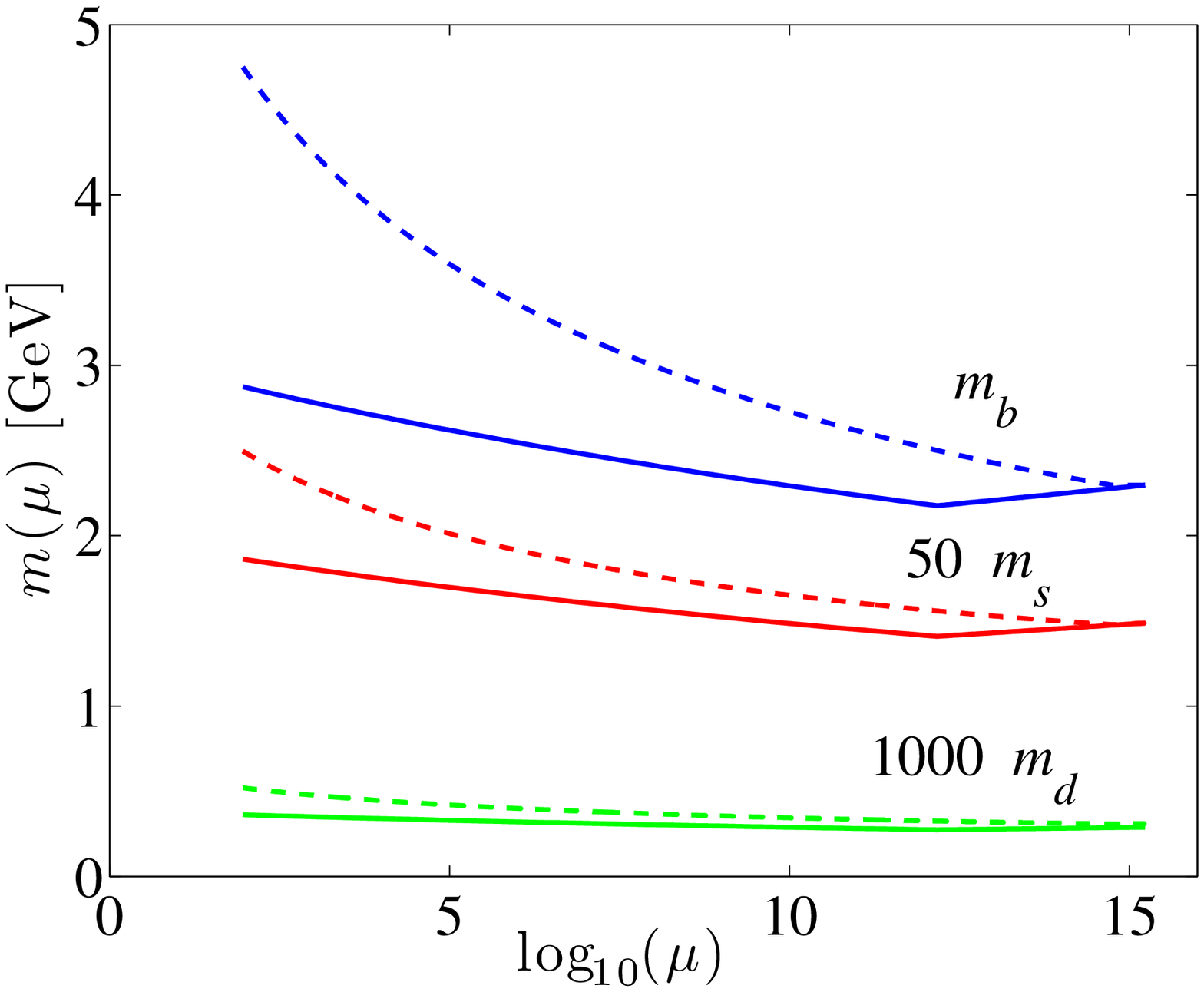}}
\caption{\it The RG running of the up-type (left plot) and down-type (right plot) quark masses, respectively, with (solid curves) and without (dashed curves) the intermediate energy scale $M_{\rm I}$ as functions of the energy scale $\mu$.}
\label{fig:quarkmasses}
\end{center}
\end{figure}

\begin{figure}
\begin{center}
\subfigure{\includegraphics[scale=0.38]{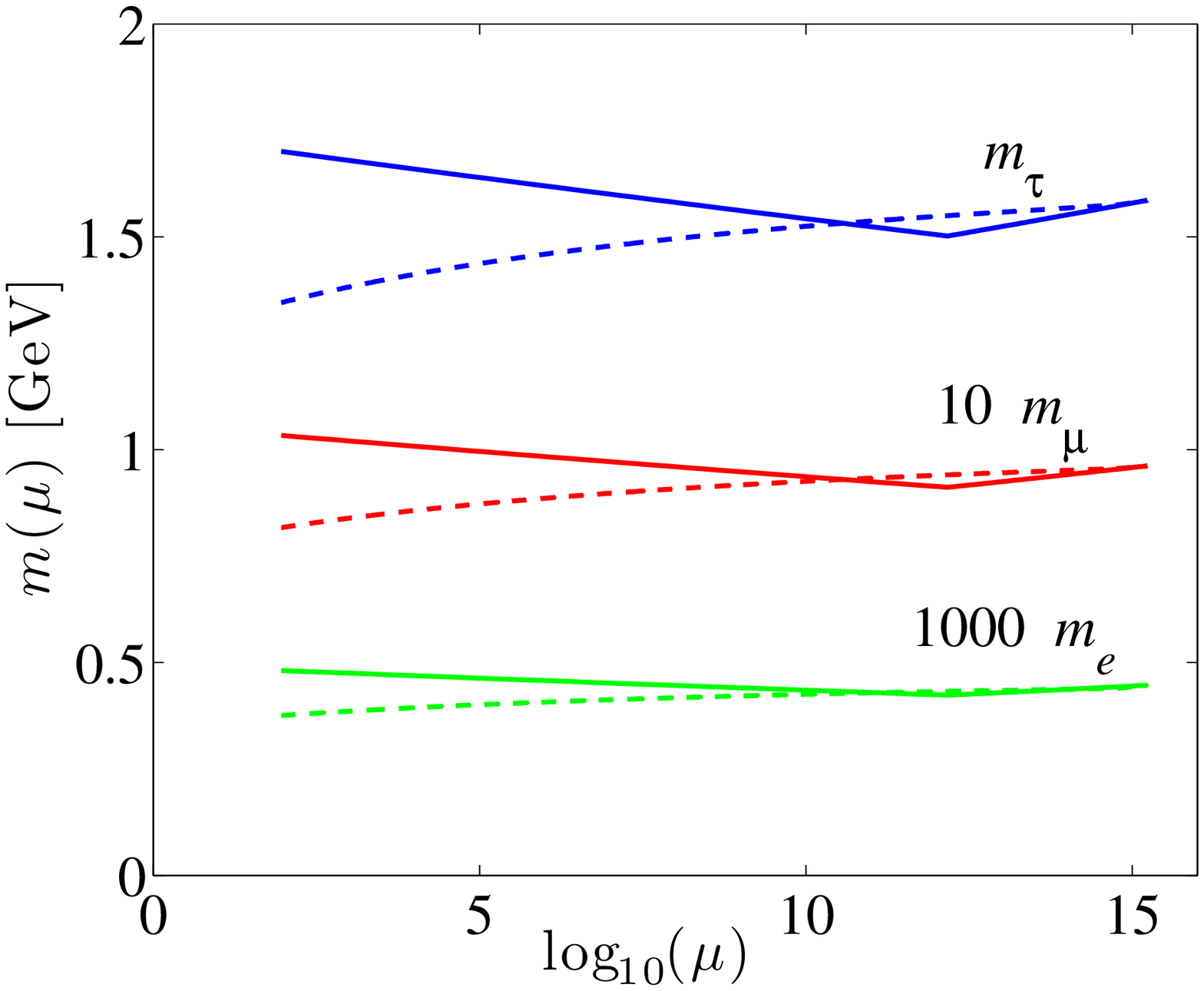}}
\subfigure{\includegraphics[scale=0.38]{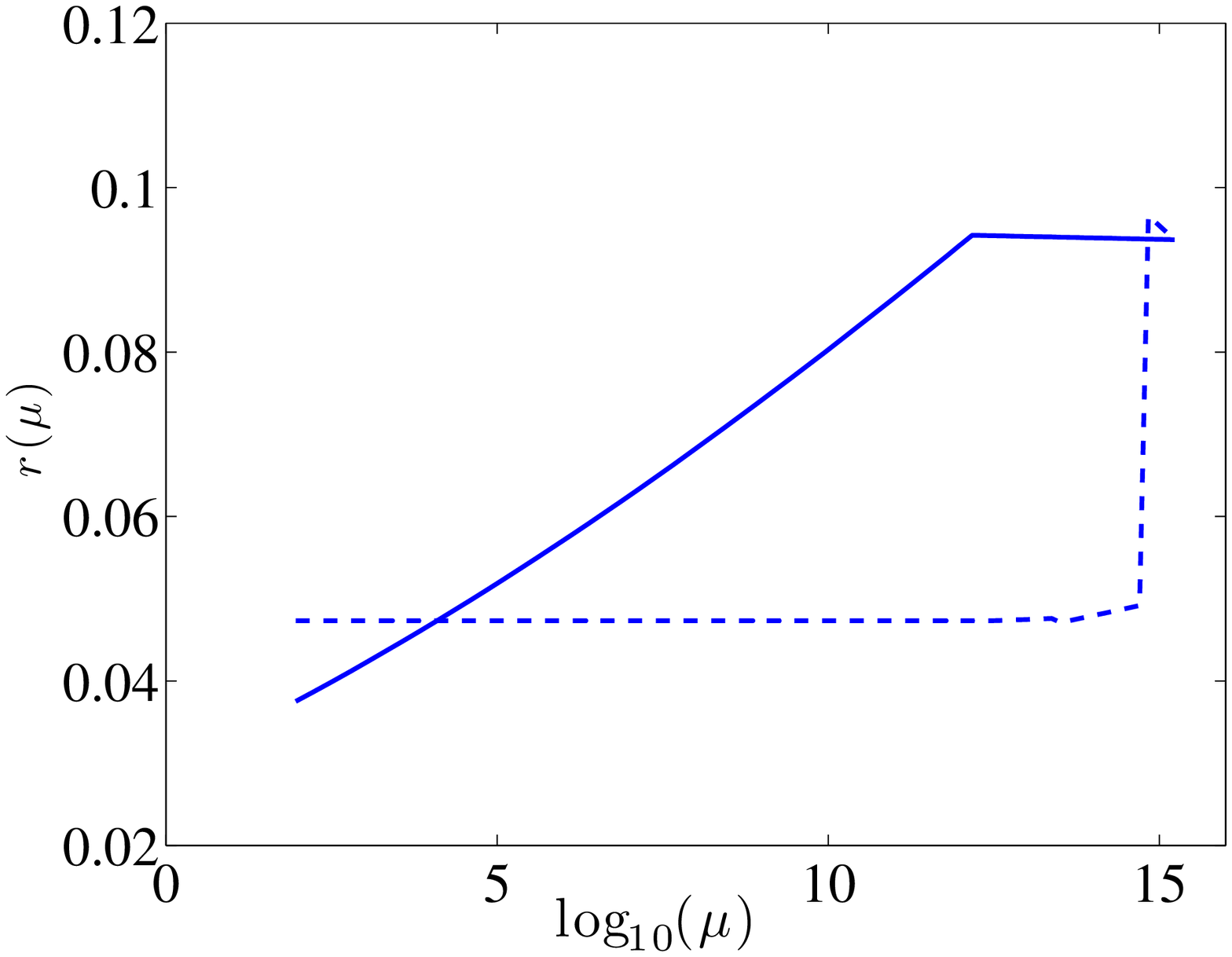}}
\caption{\it The RG running of the charged lepton masses (left plot) and the ratio of the small and large neutrino mass-squared differences (right plot), respectively, with (solid curves) and without (dashed curves) the intermediate energy scale $M_{\rm I}$ as functions of the energy scale $\mu$.}
\label{fig:leptonmasses}
\end{center}
\end{figure}

\begin{figure}
\begin{center}
\subfigure{\includegraphics[scale=0.38]{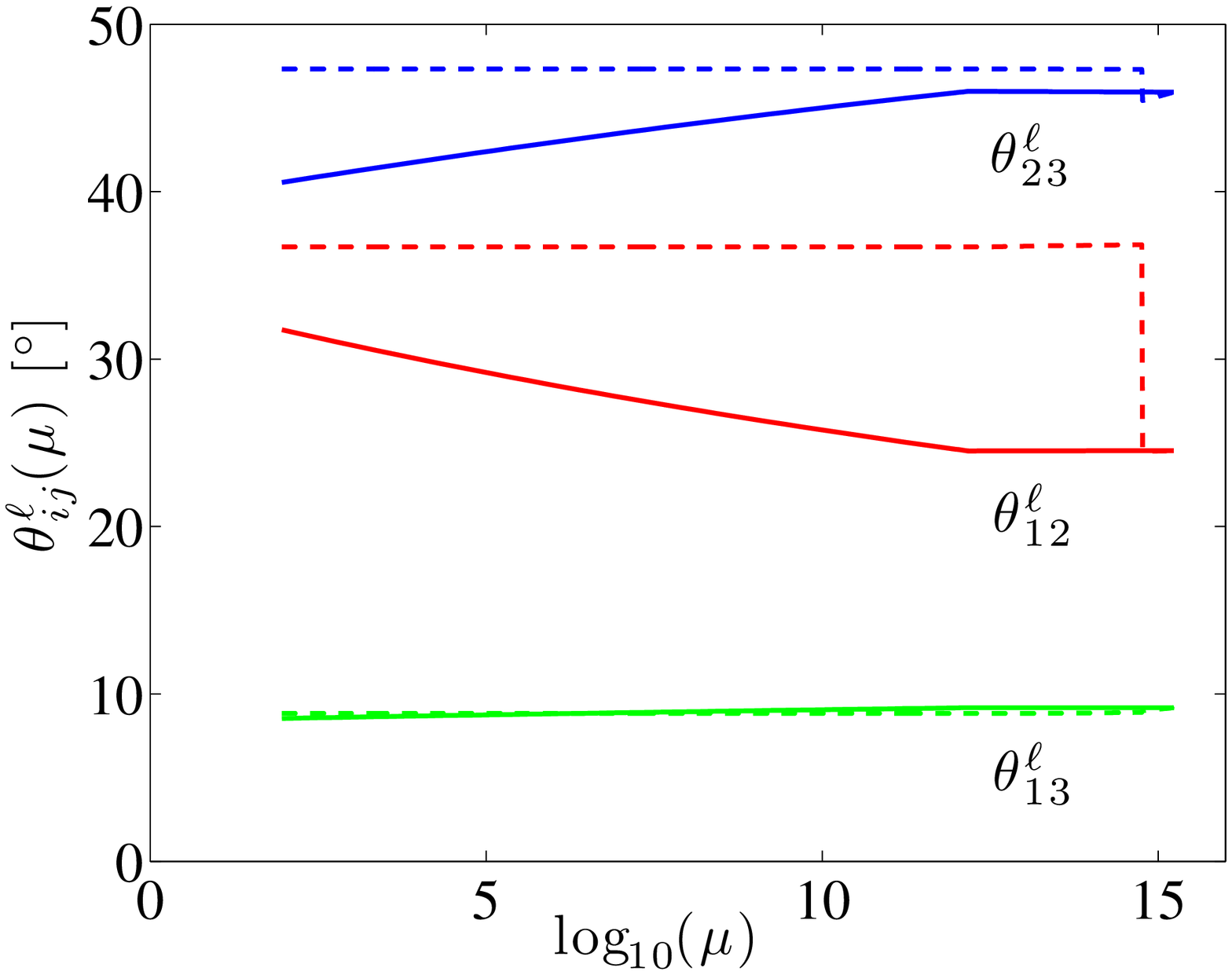}}
\subfigure{\includegraphics[scale=0.38]{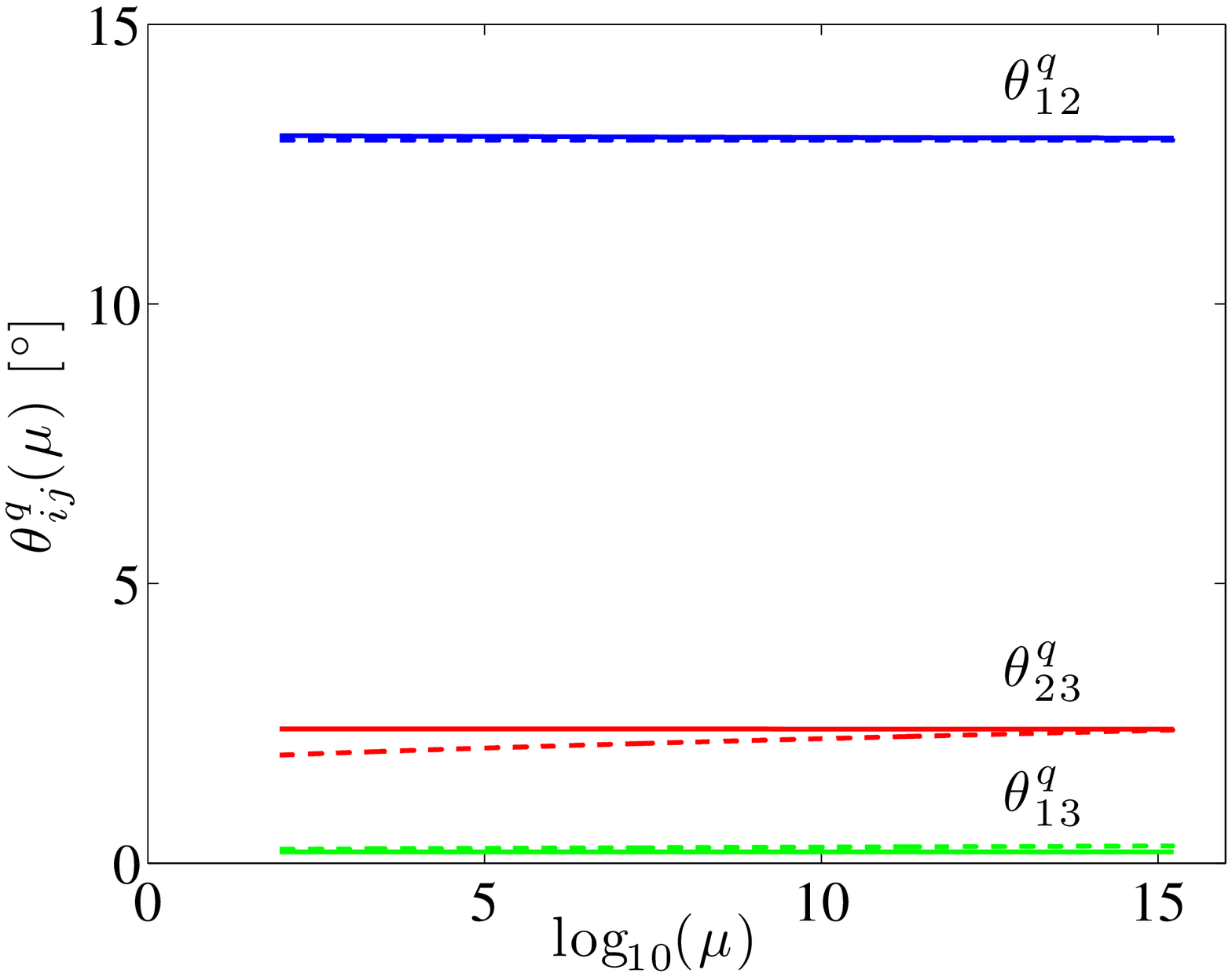}}
\caption{\it The RG running of the leptonic mixing angles (left plot) and the quark mixing parameters (right plot), respectively, with (solid curves) and without (dashed curves) the intermediate energy scale $M_{\rm I}$ as functions of the energy scale $\mu$.}
\label{fig:angleRun}
\end{center}
\end{figure}

In order to better perceive the impact of $M_{\rm I}$, we show the results of the RG running of the fermion observables from $M_{\rm GUT}$ down to $M_{\rm Z}$ (solid curves in Figs.~\ref{fig:quarkmasses}--\ref{fig:angleRun}), {\it i.e.}~the numerical solutions to the RGEs for the six quark masses (Fig.~\ref{fig:quarkmasses}), the three charged lepton masses and the ratio of the small and large neutrino mass squared differences (Fig.~\ref{fig:leptonmasses}), and the three leptonic mixing angles and the three quark mixing angles (Fig.~\ref{fig:angleRun}). These results are compared with the case where there is no intermediate scale $M_{\rm I}$, {\it i.e.}~solving the RGEs assuming the same values of $h$ and $f$ given in Eqs.~\eqref{eq:h} and \eqref{eq:f} and performing the RG running from $M_{\rm GUT}$ down to $M_{\rm Z}$ (dashed curves in Figs.~\ref{fig:quarkmasses}--\ref{fig:angleRun}). The model we use for this comparison is the SM with a type-I seesaw in which the three heavy neutrinos are integrated out at different energy scales. 
For the RG running in the SM-like model, we use the same starting point at $M_{\rm GUT}$ [as in the case of the SO(10) model] 
in order to quantify the impact of $M_{\rm I}$ at $M_{\rm Z}$. 
Note that one can compare the two models in two different ways. 
In the first case, one can use the same starting point at $M_{\rm GUT}$, then evolve the 
two models down to $M_{\rm Z}$ and there make a comparison. In the second case, one can make a new fit using the SM RGEs, 
which would then reproduce the experimental values at $M_{\rm Z}$ and then compare the two models at $M_{\rm GUT}$. 
In the present analysis, we have chosen to use the first case for the comparison. The RG running for the SM-like model was performed using the Mathematica software package REAP \cite{Antusch:2005gp}.
% without $M_{\rm I}$, 
% a significant difference was obtained for the RG 
% running of the quark and charged lepton masses. This difference is illustrated in Figs.~\ref{fig:quarkmasses}--\ref{fig:angleRun}, 
% where the RG running of the masses with $M_{\rm I}$ is drawn with solid curves, and the RG running without $M_{\rm I}$ is given by 
% the dashed curves. However, note that $M_{\rm I}$ does not have any significant effect on the RG running of the neutrino 
% observables (i.e., the two mass-squared differences and the three leptonic mixing angles). Then, it is rather the seesaw scale, 
% which still is taken into consideration, that is of importance for the evolution. The RG running of the neutrino mass observables 
% $\Delta m_{31}^2$ and $\Delta m_{21}^2$ are given in the right plot of Fig.~\ref{fig:leptonmasses}, whereas the RG running of the 
% leptonic mixing angles $\theta^\ell_{12}$, $\theta^\ell_{13}$, and $\theta^\ell_{23}$ are displayed in Fig.~\ref{fig:angleRun}. 

Now, we will discuss the results presented in Figs.~\ref{fig:quarkmasses}--\ref{fig:angleRun} in some more depth and detail. First, in Fig.~\ref{fig:quarkmasses}, in the case of the model with $M_{\rm I}$, we observe that the 
slope of the RG running of the quark masses changes direction at $M_{\rm I}$: 
from $M_{\rm GUT}$ down to $M_{\rm I}$, it decreases monotonically, whereas from $M_{\rm I}$ down to $M_{\rm Z}$, 
it increases monotonically. The reason for this change of direction in the evolution can be deduced from the change of sign 
in front of the gauge coupling terms, which dominate 
the $\beta$-functions in the RGEs that are given in Appendices \ref{app:A1} and \ref{app:A2}. As expected, in the case of the model without $M_{\rm I}$, {\it i.e.}~the SM case, the RG running 
from $M_{\rm GUT}$ down to $M_{\rm Z}$ increases monotonically.
Thus, at the $M_{\rm Z}$, the quark masses in the two cases will differ, and they will be larger in the SM case than in 
the model with $M_{\rm I}$. The smallest difference is for the top quark mass, which is 10~\% larger at $M_{\rm Z}$, 
whereas the largest difference is for the bottom quark mass, which is 65~\% larger. The other differences at $M_{\rm Z}$ 
are 21~\%, 44~\%, 36~\%, and 34~\% for the up, down, charm, and strange quark, respectively. 
In general, the relative RG running for the quark masses is substantial, both with and without $M_{\rm I}$ and it is 
essentially of the same size for both the up-type and down-type quarks.

Then, in Fig.~\ref{fig:leptonmasses}, the RG running of the lepton masses is presented. 
In the left plot of Fig.~\ref{fig:leptonmasses}, we display the RG running of the charged lepton masses, which 
exhibits a similar pattern to that of the RG running of the quark masses for the model with $M_{\rm I}$. 
However, in the SM case, the RG running has the opposite direction, {\it i.e.}~it decreases 
monotonically from $M_{\rm GUT}$ to $M_{\rm Z}$. Hence, at $M_{\rm Z}$, all charged lepton masses are approximately 21~\% smaller 
in the SM case than in the model with $M_{\rm I}$. Note that there is no obvious cause for the decrease of the charged lepton 
masses and the increase of the quark masses from the RGEs, which are given in Ref.~\cite{Antusch:2005gp}, but rather a combined 
effect of several different terms in these equations. In the right plot of Fig.~\ref{fig:leptonmasses}, we show the RG running 
of the quantity $r \equiv \Delta m_{21}^2/\Delta m_{31}^2$. This quantity exhibits significant RG running in both models, 
even though the behavior is rather different. However, this is to be expected, since the models differ most significantly 
in the neutrino sector. In the so-called SM, there are three seesaw scales which have a large effect on the RG running of $r$. In particular, the most substantial effect is caused by the crossing of the threshold imposed by the largest heavy neutrino mass, 
which is around $5.8\cdot 10^{14}$~GeV.
The RG running in the model with $M_{\rm I}$ is moderate from $M_{\rm GUT}$ down to $M_{\rm I}$ but significant from $M_{\rm I}$ 
down to $M_{\rm Z}$. At $M_{\rm Z}$, the value of $r$ is 27~\% larger in the SM than in the model with $M_{\rm I}$. 

Finally, in Fig.~\ref{fig:angleRun}, we present the RG running of the leptonic and quark mixing angles. 
In the left plot of Fig.~\ref{fig:angleRun}, we display the RG running of the leptonic mixing angles. 
The evolution of these angles is negligible between $M_{\rm GUT}$ and the seesaw scale for the model with $M_{\rm I}$. 
However, from the seesaw scale down to $M_{\rm Z}$, $\theta^\ell_{12}$ increases monotonically whereas $\theta^\ell_{23}$ 
decreases monotonically. The RG running of $\theta^\ell_{13}$ is negligible. 
The exact reason for the behavior of the RG 
running of each parameter is difficult to pinpoint, since we evolve the Yukawa matrices and not the leptonic mixing angles themselves. However, the magnitude of the 
RG running is what would be expected from other analyses of seesaw models (see, e.g., Ref.~\cite{Ohlsson:2013xva} and references therein). In the SM, the angles are all larger at $M_{\rm Z}$, with the smallest difference occurring for 
$\theta^\ell_{13}$, which is only 3~\% larger, and the largest difference for $\theta^\ell_{23}$, which is 17~\% larger. 
The difference for $\theta^\ell_{23}$ is 16~\% larger. In the right plot of Fig.~\ref{fig:angleRun}, we show the RG running of the quark mixing angles. Unlike the other observables, we do not see a significant impact of 
$M_{\rm I}$ on the evolution of the quark
mixing angles, except for $\theta^q_{23}$, which, in the SM, is 20~\% smaller at $M_{\rm Z}$.
To conclude, the RG running in the model with $M_{\rm I}$ is naturally rather different from previous models presented in the 
literature, which is clearly realized in the comparison with the SM.

\section{Summary and conclusions}
\label{summary}

In this work, we have explored the effects of an intermediate energy scale on the evolution of the fermion masses and mixings in an SO(10) model with a Pati--Salam intermediate gauge group. The effects have been compared to the evolution from the GUT scale down to the EW scale in a SM-like model with three additional right-handed neutrinos. In order to quantify the differences between the two models, we have first determined the entries of the Yukawa couplings $h$ and $f$ at the GUT scale, such that the fermion observables at the EW scale are reproduced with good accuracy in this SO(10) model. The same values of the $h$ and $f$ couplings were then used as a starting point at the GUT scale for the RG running in the SM, which allows for a comparison at the EW scale. We have found that the solutions to the RGEs, {\it i.e.}, the values of the fermion observables, at the EW scale in the SM, disagree compared to the SO(10) model well beyond experimental uncertainties, which are at the level of 30~\% for the quark masses. Note that there is basically no RG running of the quark mixing angles, neither in the SO(10) model with an intermediate energy scale nor in the SM-like model. Thus, the result of our analysis is that the presence of intermediate scales has significant effects on the RG running of the fermion observables, and therefore, such intermediate scales must be taken into account in computations for a GUT model with intermediate gauge groups.

\section*{Acknowledgments}

We acknowledge the hospitality and support from the NORDITA scientific program ``News in Neutrino Physics'', April 7--May 2, 2014 during which parts of this study were performed. We would like to thank 
K.S.~Babu, Johannes Bergstr\"{o}m, and He Zhang for useful discussions.

This work was supported by MIUR (Italy) under the program Futuro in Ricerca 2010 (RBFR10O36O) (D.M.) and the Swedish Research Council (Vetenskapsr{\aa}det), contract no. 621-2011-3985 (T.O.).

\appendix

\section{Renormalization group equations}
\label{app:A}

In this appendix, we list the RGEs for (i) the Yukawa couplings from $M_{\rm GUT}$ to $M_{\rm I}$, (ii) the Yukawa couplings from $M_{\rm I}$ to $M_{\rm Z}$, and (iii) RGEs for the effective neutrino mass matrix.

\subsection{RGEs for the Yukawa couplings from $\mbox{\boldmath$M_{\rm GUT}$}$ to $\mbox{\boldmath$M_{\rm I}$}$}
\label{app:A1}

Firstly, we present the RGEs for the Yukawa couplings from the GUT scale $M_{\rm GUT}$ to the intermediate scale $M_{\rm I}$, which are given by
\bea
16\pi^2\frac{\dr Y^{(10)}_F}{\dr t}&=&\left(Y^{(10)}_FY^{(10)\dagger}_F
+\frac{15}{4}Y^{(126)}_FY^{(126)\dagger}_F\right)Y^{(10)}_F \nonumber\\
&+&Y^{(10)}_F\left\{Y^{(10)}_FY^{(10)\dagger}_F
+\frac{15}{4}\left(Y^{(126)}_F Y^{(126)\dagger}_F
+Y^{(126)}_R Y^{(126)\dagger}_R \right)\right\} \nonumber\\
&+&4{\mathrm{tr}}\left(Y^{(10)}_F Y^{(10)\dagger}_F\right)Y^{(10)}_F
+\left(\frac{9}{4}g_{2L}^{2}+\frac{9}{4}g_{2R}^{2}
+\frac{15}{4}g_{4C}^{\,2}\right) Y^{(10)}_F\,,
\eea
\bea
16\pi^2\frac{\dr Y^{(126)}_F}{\dr t}&=&\left(Y^{(10)}_FY^{(10)\dagger}_F
+\frac{15}{4}Y^{(126)}_FY^{(126)\dagger}_F\right)Y^{(126)}_F \nonumber\\
&+&Y^{(126)}_F\left\{Y^{(10)}_FY^{(10)\dagger}_F
+\frac{15}{4}\left(Y^{(126)}_FY^{(126)\dagger}_F
+Y^{(126)}_R Y^{(126)\dagger}_R \right)\right\} \nonumber\\
&+&{\mathrm{tr}}\left(Y^{(126)}_FY^{(126)\dagger}_F\right)Y^{(126)}_F
+\left(\frac{9}{4}g_{2L}^{2}+\frac{9}{4}g_{2R}^{2}
+\frac{15}{4}g_{4C}^{2}\right) Y^{(126)}_F\,,
\eea
\bea
16\pi^2\frac{\dr Y^{(126)}_R}{\dr t}&=&\left\{Y^{(10)}_FY^{(10)\dagger}_F
+\frac{15}{4}\left(Y^{(126)}_FY^{(126)\dagger}_F
+Y^{(126)}_R Y^{(126)\dagger}_R \right)\right\}Y^{(126)}_R \nonumber\\
&+&Y^{(126)}_R\left\{Y^{(10)}_FY^{(10)\dagger}_F
+\frac{15}{4}\left(Y^{(126)}_FY^{(126)\dagger}_F
+Y^{(126)}_R Y^{(126)\dagger}_R \right)\right\} \nonumber\\
&+&{\mathrm{tr}}\left(Y^{(126)}_RY^{(126)\dagger}_R\right)Y^{(126)}_R
+\left(\frac{9}{2}g_{2R}^{2}
+\frac{15}{4}g_{4C}^{2}\right) Y^{(126)}_R\,,
\eea
where $g_{\rm 2L}$, $g_{\rm 2R}$, and $g_{\rm 4C}$ are the ${\rm SU(2)}_{\rm L}$, ${\rm SU(2)}_{\rm R}$, and ${\rm SU(4)}_{\rm C}$ gauge coupling constants, respectively.

\subsection{RGEs for the Yukawa couplings from $\mbox{\boldmath$M_{\rm I}$}$ to $\mbox{\boldmath$M_{\rm Z}$}$}
\label{app:A2}

Secondly, we present the RGEs for the Yukawa couplings from the intermediate scale $M_{\rm I}$ to the electroweak scale $M_{\rm Z}$, which are given by
\bea
16\pi^2\frac{\dr Y^{(10)}_u}{\dr t}&=&3{\mathrm{tr}}
(Y^{(10)}_uY^{(10)\dagger}_u)Y^{(10)}_u
+3{\mathrm{tr}}(Y^{(10)}_uY^{(126)\dagger}_u)Y^{(126)}_u \nonumber\\
&-&\left(8g_3^2+\frac{9}{4}g_2^2
+\frac{17}{12}g_Y^2\right)Y_u^{(10)} \nonumber\\
&+&\frac{1}{2}\big(Y^{(10)}_uY^{(10)\dagger}_u 
+Y^{(126)}_uY^{(126)\dagger}_u \nonumber\\
&+&Y^{(10)}_dY^{(10)\dagger}_d
+Y^{(126)}_dY^{(126)\dagger}_d \big)Y^{(10)}_u \nonumber\\
&+&Y^{(10)}_u\left(Y^{(10)\dagger}_uY^{(10)}_u
 +Y^{(126)\dagger}_uY^{(126)}_u\right)\,,\label{yu10} 
\eea
\bea
16\pi^2\frac{\dr Y^{(126)}_u}{\dr t}&=&3{\mathrm{tr}}
(Y^{(126)}_uY^{(126)\dagger}_u)Y^{(126)}_u
+3{\mathrm{tr}}(Y^{(126)}_uY^{(10)\dagger}_u)Y^{(10)}_u \nonumber\\
&-&\left(8g_3^2+\frac{9}{4}g_2^2+\frac{17}{12}g_Y^2\right)
Y_u^{(126)} \nonumber\\
&+&\frac{1}{2}\big(Y^{(10)}_uY^{(10)\dagger}_u
+Y^{(126)}_uY^{(126)\dagger}_u \nonumber\\
&+&Y^{(10)}_dY^{(10)\dagger}_d
+Y^{(126)}_dY^{(126)\dagger}_d \big)Y^{(126)}_u \nonumber\\
&+&Y^{(126)}_u\left(Y^{(10)\dagger}_uY^{(10)}_u 
+Y^{(126)\dagger}_uY^{(126)}_u\right)\,,
\eea
\bea
16\pi^2\frac{\dr Y^{(10)}_d}{\dr t}&=&\left\{3{\mathrm{tr}}
(Y^{(10)}_dY^{(10)\dagger}_d)+{\mathrm{tr}}
(Y^{(10)}_eY^{(10)\dagger}_e)\right\}Y^{(10)}_d \nonumber\\
&+&\left\{3{\mathrm{tr}}(Y^{(10)}_dY^{(126)\dagger}_d)
+{\mathrm{tr}}(Y^{(10)}_eY^{(126)\dagger}_e)\right\}Y^{(126)}_d \nonumber\\
&-&\left(8g_3^2+\frac{9}{4}g_2^2
+\frac{5}{12}g_Y^2\right)Y_d^{(10)} \nonumber\\
&+&\frac{1}{2}\Big(Y^{(10)}_uY^{(10)\dagger}_u 
+ Y^{(126)}_uY^{(126)\dagger}_u \nonumber\\
&+&Y^{(10)}_dY^{(10)\dagger}_d 
+Y^{(126)}_dY^{(126)\dagger}_d \Big)Y^{(10)}_d  \nonumber\\
&+&Y^{(10)}_d\left(Y^{(10)\dagger}_dY^{(10)}_d 
+Y^{(126)\dagger}_dY^{(126)}_d\right)\,,
\eea
\bea
16\pi^2\frac{\dr Y^{(126)}_d}{\dr t}&=&\left\{3{\mathrm{tr}}
(Y^{(126)}_dY^{(126)\dagger}_d)+{\mathrm{tr}}
(Y^{(126)}_eY^{(126)\dagger}_e)\right\}Y^{(126)}_d \nonumber\\
&+&\left\{3{\mathrm{tr}}(Y^{(126)}_dY^{(10)\dagger}_d)
+{\mathrm{tr}}(Y^{(126)}_eY^{(10)\dagger}_e)\right\}Y^{(10)}_d \nonumber\\
&-&\left(8g_3^2+\frac{9}{4}g_2^2
+\frac{5}{12}g_Y^2\right)Y_d^{(126)} \nonumber\\
&+&\frac{1}{2}\Big(Y^{(10)}_uY^{(10)\dagger}_u 
+Y^{(126)}_uY^{(126)\dagger}_u \nonumber\\
&+&Y^{(10)}_dY^{(10)\dagger}_d 
+Y^{(126)}_dY^{(126)\dagger}_d \Big)Y^{(126)}_d  \nonumber\\
&+&Y^{(126)}_d\left(Y^{(10)\dagger}_dY^{(10)}_d 
+Y^{(126)\dagger}_dY^{(126)}_d\right)\,,
\eea
\bea
16\pi^2\frac{\dr Y^{(10)}_e}{\dr t}&=&\left\{3{\mathrm{tr}}
(Y^{(10)}_dY^{(10)\dagger}_d)+{\mathrm{tr}}
(Y^{(10)}_eY^{(10)\dagger}_e)\right\}Y^{(10)}_e \nonumber\\
&+&\left\{3{\mathrm{tr}}(Y^{(10)}_dY^{(126)\dagger}_d)
+{\mathrm{tr}}(Y^{(10)}_eY^{(126)\dagger}_e)\right\}Y^{(126)}_e \nonumber\\
&-&\left(\frac{9}{4}g_2^2+\frac{15}{4}g_Y^2\right)Y_e^{(10)} \nonumber\\
&+&\frac{1}{2}\left(Y^{(10)}_eY^{(10)\dagger}_e 
+Y^{(126)}_eY^{(126)\dagger}_e \right)Y^{(10)}_e \nonumber\\
&+&Y^{(10)}_e\left(Y^{(10)\dagger}_eY^{(10)}_e
+Y^{(126)\dagger}_eY^{(126)}_e\right)\,,
\eea
\bea
16\pi^2\frac{\dr Y^{(126)}_e}{\dr t}&=&\left\{3{\mathrm{tr}}
(Y^{(126)}_dY^{(126)\dagger}_d)+{\mathrm{tr}}
(Y^{(126)}_eY^{(126)\dagger}_e)\right\}Y^{(126)}_e \nonumber\\
&+&\left\{3{\mathrm{tr}}(Y^{(126)}_dY^{(10)\dagger}_d)
+{\mathrm{tr}}(Y^{(126)}_eY^{(10)\dagger}_e)\right\}Y^{(10)}_e \nonumber\\ 
&-&\left(\frac{9}{4}g_2^2+\frac{15}{4}g_Y^2\right)
Y_e^{(126)} \nonumber\\
&+&\frac{1}{2}\left(Y^{(10)}_eY^{(10)\dagger}_e
+Y^{(126)}_eY^{(126)\dagger}_e \right)Y^{(126)}_e \nonumber\\
&+&Y^{(126)}_e\left(Y^{(10)\dagger}_eY^{(10)}_e
+Y^{(126)\dagger}_eY^{(126)}_e\right)\,,\label{ye126}
\eea
where $g_3$, $g_2$, and $g_{\rm Y}$ are the ${\rm SU(3)}_{\rm C}$, ${\rm SU(2)}_{\rm L}$, and ${\rm U(1)}_{\rm Y}$ gauge coupling constants, respectively.

\subsection{RGEs for the effective neutrino mass matrix}
\label{app:A3}

Similarly, we display the RGEs for coefficients of the effective neutrino mass matrix, which are given by
\begin{eqnarray}
16\pi^2\frac{\dr\kappa^{(1,1)}}{\dr t}&=&6{\mathrm{tr}}
\left(Y^{(10)}_uY^{(10)\dagger}_u\right)\kappa^{(1,1)} \nonumber\\
&+&3{\mathrm{tr}}\left(Y^{(126)}_uY^{(10)\dagger}_u\right)
\left(\kappa^{(1,2)}+\kappa^{(2,1)}\right) \nonumber\\
&-&3g_2^2\kappa^{(1,1)} \nonumber\\
&+&\frac{1}{6}\left(\lambda_{1111}\kappa^{(1,1)}+\lambda_{1212}\kappa^{(2,2)}\right) \nonumber\\
&+&\frac{1}{2}\Big\{\left(Y^{(10)}_eY^{(10)\dagger}_e
+Y^{(126)}_eY^{(126)\dagger}_e\right)\kappa^{(1,1)} \nonumber\\
&+&\kappa^{(1,1)}\left(Y^{(10)}_eY^{(10)\dagger}_e
+Y^{(126)}_eY^{(126)\dagger}_e\right)^T\Big\}\,,
\eea
\bea
16\pi^2\frac{\dr\kappa^{(2,2)}}{\dr t}&=&6{\mathrm{tr}}
\left(Y^{(126)}_uY^{(126)\dagger}_u\right)\kappa^{(2,2)} \nonumber\\
&+&3{\mathrm{tr}}\left(Y^{(10)}_uY^{(126)\dagger}_u\right)
\left(\kappa^{(1,2)}+\kappa^{(2,1)}\right) \nonumber\\
&-&3g_2^2\kappa^{(2,2)} \nonumber\\
&+&\frac{1}{6}\left(\lambda_{2222}\kappa^{(2,2)}
+\lambda_{2121}\kappa^{(1,1)}\right) \nonumber\\
&+&\frac{1}{2}\Big\{\left(Y^{(10)}_eY^{(10)\dagger}_e
+Y^{(126)}_eY^{(126)\dagger}_e\right)\kappa^{(2,2)} \nonumber\\
&+&\kappa^{(2,2)}\left(Y^{(10)}_eY^{(10)\dagger}_e
+Y^{(126)}_eY^{(126)\dagger}_e\right)^T\Big\}\,,
\eea
\bea
16\pi^2\frac{\dr\kappa^{(1,2)}}{\dr t}&=&3{\mathrm{tr}}
\left(Y^{(10)}_uY^{(10)\dagger}_u
+Y^{(126)}_uY^{(126)\dagger}_u\right)\kappa^{(1,2)} \nonumber\\
&+&3{\mathrm{tr}}\left(Y^{(10)}_uY^{(126)\dagger}_u\right)
\kappa^{(1.1)} \nonumber\\
&+&3{\mathrm{tr}}\left(Y^{(126)}_uY^{(10)\dagger}_u\right)
\kappa^{(2,2)} \nonumber\\
&-&g_2^2\left(2\kappa^{(1,2)}+\kappa^{(2,1)}\right) \nonumber\\
&+&\frac{1}{6}\left(\lambda_{1122}\kappa^{(1,2)}
+\lambda_{1221}\kappa^{(2,1)}\right) \nonumber\\
&+&\frac{1}{2}\Big\{\left(Y^{(10)}_eY^{(10)\dagger}_e
+Y^{(126)}_eY^{(126)\dagger}_e\right)\kappa^{(1,2)} \nonumber\\
&+&\kappa^{(1,2)}\left(Y^{(10)}_eY^{(10)\dagger}_e
+Y^{(126)}_eY^{(126)\dagger}_e\right)^T\Big\}\,,
\eea
\bea
16\pi^2\frac{\dr \kappa^{(2,1)}}{\dr t}&=&3{\mathrm{tr}}
\left(Y^{(10)}_uY^{(10)\dagger}_u
+Y^{(126)}_uY^{(126)\dagger}_u\right)\kappa^{(2,1)} \nonumber\\
&+&3{\mathrm{tr}}\left(Y^{(10)}_uY^{(126)\dagger}_u\right)
\kappa^{(1.1)} \nonumber\\
&+&3{\mathrm{tr}}\left(Y^{(126)}_uY^{(10)\dagger}_u\right)
\kappa^{(2,2)} \nonumber\\
&-&g_2^2\left(\kappa^{(1,2)}+2\kappa^{(2,1)}\right) \nonumber\\
&+&\frac{1}{6}\left(\lambda_{2211}\kappa^{(2,1)}
+\lambda_{2112}\kappa^{(1,2)}\right) \nonumber\\
&+&\frac{1}{2}\Big\{\left(Y^{(10)}_eY^{(10)\dagger}_e
+Y^{(126)}_eY^{(126)\dagger}_e\right)\kappa^{(2,1)} \nonumber\\
&+&\kappa^{(2,1)}\left(Y^{(10)}_eY^{(10)\dagger}_e
+Y^{(126)}_eY^{(126)\dagger}_e\right)^T\Big\}\,,
\end{eqnarray}
where the parameters $\lambda_{ijlm}$ are the Higgs self-couplings, which have to be accounted for in a consistent way.

\providecommand{\href}[2]{#2}\begingroup\raggedright\endgroup


\begin{thebibliography}{10}

\bibitem{Bajc:2005zf}
B.~Bajc, A.~Melfo, G.~Senjanovi{\'c}, and F.~Vissani,  Phys. Rev. {\bf D73}
  (2006) 055001, [\href{http://xxx.lanl.gov/abs/hep-ph/0510139}{{\tt
  hep-ph/0510139}}].

\bibitem{Altarelli:2013aqa}
G.~Altarelli and D.~Meloni,  JHEP {\bf 1308} (2013) 021,
  [\href{http://xxx.lanl.gov/abs/1305.1001}{{\tt arXiv:1305.1001}}].

\bibitem{delAguila:1980at}
F.~del Aguila and L.~E. Ib{\'a}{\~n}ez,  Nucl. Phys. {\bf B177} (1981) 60--86.

\bibitem{Harvey:1981hk}
J.~A. Harvey, D.~Reiss, and P.~Ramond,  Nucl. Phys. {\bf B199} (1982) 223--268.

\bibitem{Robinett:1982tg}
R.~W. Robinett and J.~L. Rosner, Phys. Rev. {\bf D26} (1982) 2396--2419.

\bibitem{Mohapatra:1982tc}
R.~N. Mohapatra and G.~Senjanovi{\'c},  Z. Phys. {\bf C17} (1983) 53--56.

\bibitem{Babu:1992ia}
K.~Babu and R.~Mohapatra,  Phys. Rev. Lett. {\bf 70} (1993) 2845--2848,
  [\href{http://xxx.lanl.gov/abs/hep-ph/9209215}{{\tt hep-ph/9209215}}].

\bibitem{Deshpande:1992au}
N.~Deshpande, E.~Keith, and P.~B. Pal,  Phys. Rev. {\bf D46} (1993) 2261--2264.


\bibitem{Matsuda:2000zp} 
K.~Matsuda, Y.~Koide and T.~Fukuyama,
Phys.\ Rev.\ D {\bf 64}, 053015 (2001)
[hep-ph/0010026].

\bibitem{Matsuda:2001bg}
K.~Matsuda, Y.~Koide, T.~Fukuyama and H.~Nishiura,
Phys.\ Rev.\ D {\bf 65} (2002) 033008
[Erratum-ibid.\ D {\bf 65} (2002) 079904]
[hep-ph/0108202].

\bibitem{Bertolini:2009qj}
S.~Bertolini, L.~Di~Luzio, and M.~Malinsk{\'y},  Phys. Rev. {\bf D80} (2009)
  015013, [\href{http://xxx.lanl.gov/abs/0903.4049}{{\tt arXiv:0903.4049}}].

\bibitem{Bertolini:2012im}
S.~Bertolini, L.~Di~Luzio, and M.~Malinsk{\'y},  Phys. Rev. {\bf D85} (2012)
  095014, [\href{http://xxx.lanl.gov/abs/1202.0807}{{\tt arXiv:1202.0807}}].

\bibitem{Buccella:2012kc}
F.~Buccella, D.~Falcone, C.~S. Fong, E.~Nardi, and G.~Ricciardi,  Phys. Rev.
  {\bf D86} (2012) 035012, [\href{http://xxx.lanl.gov/abs/1203.0829}{{\tt
  arXiv:1203.0829}}].

\bibitem{Joshipura:2011nn}
A.~S. Joshipura and K.~M. Patel,  Phys. Rev. {\bf D83} (2011) 095002,
  [\href{http://xxx.lanl.gov/abs/1102.5148}{{\tt arXiv:1102.5148}}].

\bibitem{Dueck:2013gca}
A.~Dueck and W.~Rodejohann,  JHEP {\bf 1309} (2013) 024,
  [\href{http://xxx.lanl.gov/abs/1306.4468}{{\tt arXiv:1306.4468}}].

\bibitem{PhysRevD.10.275}
J.~C. Pati and A.~Salam,  Phys. Rev. {\bf D10} (1974) 275--289.

\bibitem{Koh:1983ir}
I.~Koh and S.~Rajpoot,  Phys. Lett. {\bf B135} (1984) 397--401.

\bibitem{PhysRevD.25.581}
D.~R.~T. Jones,  Phys. Rev. {\bf D25} (1982) 581--582.

\bibitem{Amsler:2008zzb}
{\bf Particle Data Group} Collaboration, C.~Amsler {\em et.~al.},  Phys. Lett.
  {\bf B667} (2008) 1--1340.

\bibitem{Mohapatra:1992dx}
R.~Mohapatra and M.~Parida,  Phys. Rev. {\bf D47} (1993) 264--272,
  [\href{http://xxx.lanl.gov/abs/hep-ph/9204234}{{\tt hep-ph/9204234}}].

\bibitem{Fukuyama:2002vv}
T.~Fukuyama and T.~Kikuchi,  Mod. Phys. Lett. {\bf A18} (2003) 719--731,
  [\href{http://xxx.lanl.gov/abs/hep-ph/0206118}{{\tt hep-ph/0206118}}].

\bibitem{Aulakh:2002zr}
C.~S. Aulakh and A.~Girdhar,  Int. J. Mod. Phys. {\bf A20} (2005) 865--894,
  [\href{http://xxx.lanl.gov/abs/hep-ph/0204097}{{\tt hep-ph/0204097}}].

\bibitem{Dutta:2004zh}
B.~Dutta, Y.~Mimura, and R.~Mohapatra,  Phys. Rev. Lett. {\bf 94} (2005)
  091804, [\href{http://xxx.lanl.gov/abs/hep-ph/0412105}{{\tt
  hep-ph/0412105}}].

\bibitem{Dutta:2005ni}
B.~Dutta, Y.~Mimura, and R.~Mohapatra,  Phys. Rev. {\bf D72} (2005) 075009,
  [\href{http://xxx.lanl.gov/abs/hep-ph/0507319}{{\tt hep-ph/0507319}}].

\bibitem{Altarelli:2010at}
G.~Altarelli and G.~Blankenburg,  JHEP {\bf 1103} (2011) 133,
  [\href{http://xxx.lanl.gov/abs/1012.2697}{{\tt arXiv:1012.2697}}].

\bibitem{Grimus:2004yh}
W.~Grimus and L.~Lavoura,  Eur. Phys. J. {\bf C39} (2005) 219--227,
  [\href{http://xxx.lanl.gov/abs/hep-ph/0409231}{{\tt hep-ph/0409231}}].

\bibitem{Cheng:1973nv}
T.~Cheng, E.~Eichten, and L.-F. Li,  Phys. Rev. {\bf D9} (1974) 2259--2273.

\bibitem{Feroz:2007kg}
F.~Feroz and M.~Hobson,  Mon. Not. Roy. Astron. Soc. {\bf 384} (2008) 449--463,
  [\href{http://xxx.lanl.gov/abs/0704.3704}{{\tt arXiv:0704.3704}}].

\bibitem{Feroz:2008xx}
F.~Feroz, M.~Hobson, and M.~Bridges,  Mon. Not. Roy. Astron. Soc. {\bf 398}
  (2009) 1601--1614, [\href{http://xxx.lanl.gov/abs/0809.3437}{{\tt
  arXiv:0809.3437}}].

\bibitem{Feroz:2013hea}
F.~Feroz, M.~Hobson, E.~Cameron, and A.~Pettitt,
  \href{http://xxx.lanl.gov/abs/1306.2144}{{\tt arXiv:1306.2144}}.

\bibitem{Skilling:2004}
J.~Skilling,  AIP Conf. Proc. {\bf 735} (2004) 395--405.

\bibitem{Skilling:2006}
J.~Skilling,  Bayesian Analysis {\bf 1} (2006) 833--860.

\bibitem{Xing:2007fb}
Z.-z. Xing, H.~Zhang, and S.~Zhou,  Phys. Rev. {\bf D77} (2008) 113016,
  [\href{http://xxx.lanl.gov/abs/0712.1419}{{\tt arXiv:0712.1419}}].

\bibitem{GonzalezGarcia:2012sz}
M.~Gonzalez-Garcia, M.~Maltoni, J.~Salvado, and T.~Schwetz,  JHEP {\bf 1212}
  (2012) 123, [\href{http://xxx.lanl.gov/abs/1209.3023}{{\tt
  arXiv:1209.3023}}].


\bibitem{Capozzi:2013csa}
F.~Capozzi, G.~L.~Fogli, E.~Lisi, A.~Marrone, D.~Montanino, and A.~Palazzo, 
Phys. Rev. {\bf D89} (2014) 093018, [\href{http://xxx.lanl.gov/abs/1312.2878}{{\tt arXiv:1312.2878}}].

\bibitem{Forero:2014bxa}
D.~V.~Forero, M.~T{\'o}rtola, and J.~W.~F.~Valle, \href{http://xxx.lanl.gov/abs/1405.7540}{{\tt arXiv:1405.7540}}.
  
\bibitem{Gonzalez-Garcia:2014bfa}
M.~C.~Gonzalez-Garcia, M.~Maltoni, and T.~Schwetz, JHEP {\bf 1411} (2014) 052, [\href{http://xxx.lanl.gov/abs/1409.5439}{{\tt arXiv:1409.5439}}].


\bibitem{Antusch:2005gp}
S.~Antusch, J.~Kersten, M.~Lindner, M.~Ratz, and M.~A. Schmidt,  JHEP {\bf
  0503} (2005) 024, [\href{http://xxx.lanl.gov/abs/hep-ph/0501272}{{\tt
  hep-ph/0501272}}].

\bibitem{Ohlsson:2013xva}
T.~Ohlsson and S.~Zhou,  Nature Commun. {\bf 5} (2014) 5153,
  [\href{http://xxx.lanl.gov/abs/1311.3846}{{\tt arXiv:1311.3846}}].


\end{thebibliography}
\end{document}